\documentclass{jfm}

\usepackage{amsfonts}
\usepackage{amsmath}
\usepackage{graphicx}
\usepackage{caption}
\usepackage{hyperref}
\usepackage[normalem]{ulem}
\usepackage[section]{placeins}
\usepackage{bm}
\usepackage{array}
\usepackage{hhline}
\usepackage{tikz}
\usepackage[ruled]{algorithm2e}
\newcommand{\be}{\begin{equation}}
\newcommand{\ee}{\end{equation}}
\newcommand{\vct}[1]{{\mbox {\boldmath $#1$}}}

\newcommand{\thickblackdashed}
    {\raisebox{2pt}{\tikz{\draw[dashed,line width = 1pt] (0,0)--(5mm,0);}}}
\newcommand{\thickbluedashed}
    {\raisebox{2pt}{\tikz{\draw[blue,dashed,line width = 1pt] (0,0)--(5mm,0);}}}
\newcommand{\thinblackdashed}
    {\raisebox{2pt}{\tikz{\draw[dashed,line width = 0.5pt] (0,0)--(5mm,0);}}}

\newcommand{\thickblackline}
    {\raisebox{2pt}{\tikz{\draw[line width = 1pt] (0,0)--(5mm,0);}}}
\newcommand{\thickredline}
    {\raisebox{2pt}{\tikz{\draw[red,line width = 1pt] (0,0)--(5mm,0);}}}
\newcommand{\thickblueline}
    {\raisebox{2pt}{\tikz{\draw[blue,line width = 1pt] (0,0)--(5mm,0);}}}
\newcommand{\thickgreenline}
    {\raisebox{2pt}{\tikz{\draw[green,line width = 1pt] (0,0)--(5mm,0);}}}
\newcommand{\thickcyanline}
    {\raisebox{2pt}{\tikz{\draw[cyan,line width = 1pt] (0,0)--(5mm,0);}}}
\newcommand{\thickgrayline}
    {\raisebox{2pt}{\tikz{\draw[gray,line width = 1pt] (0,0)--(5mm,0);}}}
\newcommand{\ihat}{\hat{\bm \imath}}
\newcommand{\jhat}{\hat{\bm \jmath}}
\newcommand{\khat}{\hat{\bm k}}
 
\shorttitle{Lagrangian Relaxation Towards Equilibrium Wall Model}
\shortauthor{M. Fowler, T.A. Zaki and C. Meneveau}

\title{A Lagrangian relaxation towards equilibrium wall model for large eddy simulation}
\author{Mitchell Fowler, Tamer A. Zaki and Charles Meneveau}
\date{April 2020}

\begin{document}

\maketitle

\begin{abstract}
A large eddy simulation wall model is developed based on a formal interpretation of quasi-equilibrium that governs the momentum balance integrated in the wall-normal direction. The model substitutes the law-of-the-wall velocity profile for a smooth surface into the wall-normal integrated momentum balance, leading to a Lagrangian relaxation towards equilibrium (LaRTE) transport equation for the friction velocity vector ${\bm u}_\tau(x,z,t)$. This partial differential equation includes a relaxation timescale governing the rate at which the wall stress can respond to imposed fluctuations due to the inertia of the fluid layer from the wall to the wall-model height. A-priori tests based on channel flow direct numerical simulation (DNS) data show that the identified relaxation timescale ensures self-consistency with assumed quasi-equilibrium conditions. The new approach enables us to formally distinguish quasi-equilibrium from additional, non-equilibrium contributions to the wall stress. A particular model for non-equilibrium contributions is derived, motivated by laminar Stokes layer dynamics in the viscous sublayer when applying fast varying pressure gradients. The new wall modeling approach is first tested in standard equilibrium channel flow in order to document various properties of the approach. The  model is then applied in LES of channel flow  with a suddenly applied spanwise pressure gradient (SSPG). The resulting mean wall stress evolution is compared to DNS with good agreement. At the onset of the SSPG, the laminar Stokes layer develops rapidly while the LaRTE portion of the stress has a delayed response due to its inherent relaxation dynamics. Results also highlight open challenges such as modeling the response of near-wall turbulence occurring above the viscous sublayer and at timescales faster than quasi-equilibrium conditions.
\end{abstract}

\setlength{\belowcaptionskip}{0pt}

\section{Introduction}
\label{sec:introduction}
Large eddy simulation (LES) has become a popular prediction tool for unsteady turbulent flows due to its ability to resolve the less universal large scale motions while modeling the more universal small scales. However, near the wall the momentum-carrying structures scale with the viscous length. Thus, LES that resolve these structures near the wall (wall-resolved LES) incur  computational costs that are not significantly lower than those of direct numerical simulation (DNS). For recent detailed analyses of computational costs of DNS and wall-resolved LES, see 
\cite{Choi2012} and \cite{Yang2021}.
Wall-resolved LES becomes computationally intractable for high Reynolds number flows relevant for many engineering and geophysical applications. Therefore, in order for LES to be a practical flow prediction tool, the near-wall region must be modeled through  wall models.

Numerous wall models have been proposed over the years \citep[see][for reviews of LES wall modeling]{Piomelli2002, Piomelli2008, Larsson2016, Bose2018}. \cite{Larsson2016} broadly identify different classes for wall models based on the coupling between the resolved and modeled regions. The wall models used and discussed in the present work are considered to be ``wall-stress models'' in which the LES formally extends all the way to the wall and the wall stress is applied as a boundary condition to the LES. The equilibrium wall model (EQWM) is the simplest and most commonly used wall-stress model. The algebraic EQWM assumes that the velocity profile follows some known functional form. Typically one assumes the velocity satisfies the ``law of the wall'' such that $\langle u \rangle^+ = f(y^+)$ where $y$ is the wall normal coordinate and ``+'' indicates inner units non-dimensionalization with the friction velocity $u_\tau$ and the kinematic viscosity $\nu$. If the wall-model height, $y_{wm}=\Delta$, lies within the log-layer then one assumes the log-law. For rough walls, this is the approach most often followed in the geophysical literature \citep{moeng84,bou05}. Inversion of a composite profile including the viscous and buffer layers is more general for smooth wall applications, see e.g. \cite{luchini2018structure,gonzalez2018large,adler2020wall}. Algebraic equilibrium wall models usually assume the velocity profile to be valid locally and instantaneously such that the LES velocity at the wall-model height may be used to find the local friction velocity and thus the local wall stress. The log-law for smooth walls is a non-linear equation for the friction velocity and thus must be solved numerically, typically using iterations. As a more practical alternative, \cite{meneveau2020note} developed fitting functions that directly enable finding the friction velocity given the velocity at the wall-model height, also including pressure gradient and roughness effects.
 
Differential forms of the equilibrium wall model are often also used through simplified forms of the thin boundary layer equations. Over the past decade, the name ``equilibrium wall model'' often refers to the numerical solution of the wall-normal diffusion equation: $\partial_y [(\nu + \nu_T)\partial_y \langle u \rangle] = 0$ where the functional form for the eddy viscosity, $\nu_T$, is assumed to be known, typically a mixing length model. 
Algebraic EQWMs are explicit solutions or approximations to the numerically solved differential EQWMs  \citep{meneveau2020note}.

Most wall models are based on the thin boundary layer equations (TBLE) which may be considered RANS-like in nature since the local momentum-carrying scales of motion are similar or smaller than the computational LES grid. So-called ``zonal'' methods, a type of hybrid method discussed in \cite{Larsson2016}, solve these equations on a separate RANS mesh, refined in the wall-normal direction, below a user-defined RANS/LES interface height. The advantage of zonal methods is their ability to include all terms in the TBLE (such as unsteady, acceleration, and pressure gradient terms), thereby allowing for greater range of applicability. However, zonal methods 
have increased cost due to the wall-normal grid refinement which ultimately can lead to costs approaching wall resolved LES. Therefore, there is continued interest towards methods that do not require 
a separate RANS mesh. The dynamic slip wall model \citep{bose_moin_2014,bae_2019} is an alternative type of wall model that does not solve RANS-like equations or make assumptions about the flow (e.g. the law of the wall).
The approach has the benefit of making connections to the dynamic model \citep{germanoetal91} using resolved flow information near the wall. However, at increasing Reynolds numbers, it is unclear that such modeling can accurately capture subtle Reynolds number dependencies of the predicted friction drag because information regarding the viscous sublayer is unavailable in such approaches. 

\cite{chung2009large} derived a wall model based on vertically integrating the unsteady term in the TBLE. A ``law of the wall'' assumption for the velocity profile is invoked such that $\langle u \rangle = u_\tau f(y^+)$ and thus upon differentiation with time the chain rule gives a $\partial u_\tau/\partial t$ term leading to an ODE in time for the wall stress. This allows obtaining an evolution equation for the wall stress as a function of known LES quantities at the model height, where advection terms in the TBLEs are approximated by the resolved LES terms, i.e. assuming plug flow profile below the wall-model height. Inspired by this work but allowing for deviations from plug flow, the integral wall model (iWMLES) introduced by \cite{yang2015integral} aims to achieve an algebraic closure based on an approach similar to the von-Karman Pohlhausen method. The assumed velocity profile is linear in the viscous sublayer up to the buffer layer and then switches  to a log profile plus a linear correction to represent possible pressure gradient effects. A series of coefficients are determined using matching conditions consisting of the boundary conditions and the full vertically-integrated TBLEs. Since iWMLES is based on the full TBLEs it can in principle capture deviations from full equilibrium conditions while still maintaining affordability due to its algebraic nature. However, while simpler than solving additional ODEs on finer meshes, the iWMLES approach still involves solving several coupled nonlinear equations, requires specifying 
an empirically chosen timescale for exponential time filtering of the velocity input at the wall-model height, and involves a discontinuity in the slope of the assumed velocity profile in the buffer region.


Additionally, in practice the so-called log-layer mismatch is a known problem in wall modeling. \cite{kawai2012wall} argue that the near-wall region of WMLES is inherently under resolved numerically, so the wall-model height should ideally be chosen further away from the wall in order to reduce the log-layer mismatch. \cite{bou05} showed through the Schwartz inequality that a local law of the wall formulation inherently over-predicts the wall stress due to LES velocity fluctuations. They then proposed reducing this error by filtering the velocity using a $2\Delta$ spatial filter. Similarly, \cite{Yang2017logmismatch} suggest local temporal filtering and/or wall-parallel spatial filtering to reduce log-layer mismatch. Recently \cite{Hosseinzade2021tfilt} tested various exponential filtering timescales of the input velocity and pressure gradient in the context of solving unsteady RANS equations with a finely resolved wall-normal grid. Unsteady and horizontal advection terms were also included. Moreover, and consistent with the recommendations by \cite{kawai2012wall}, \cite{Hosseinzade2021tfilt} found that placing the wall-model grid point inside the LES region (they used the 5th LES grid point) improved results. They also find that the use of time filtering is more important when placing the wall-model height at the first grid point as opposed to inside the LES region.

In the present work we aim to extend the iWMLES and \cite{chung2009large} approaches to arrive at a unified formulation that is both based on formal derivations starting from the underlying governing equations and is robust for applications. It will be shown that the model takes the form of a Lagrangian relaxation, hence it is termed the {\it Lagrangian relaxation towards equilibrium} (LaRTE) model. As will be seen, a number of arbitrary choices that had to be made in the context of iWMLES, such as specifying an empirically chosen exponential filtering timescale \citep[also needed in the approach by][]{Hosseinzade2021tfilt}, will no longer be required. The derivation of the LaRTE wall model is presented in \S \ref{sec:LaRTEmodel} and is based on a self-consistent interpretation of ``quasi-equilibrium'' in the assumed velocity profile below the wall-model height. This new interpretation then enables us to formally distinguish quasi-equilibrium from additional, non-equilibrium contributions to the wall stress and the latter can then be modeled separately. 

A second part of this work then concerns a model for the non-equilibrium flow and stress response in the viscous sublayer to temporally changing  pressure gradients. For applications involving time-varying applied pressure gradients, it is important to supply the new quasi-equilibrium LaRTE model with additional components that reflect the deviations from the assumed simple velocity profile in the near-wall region. Numerous prior efforts have been made to understand unsteady effects on wall-bounded turbulent flows (e.g. the works by \cite{jung_mangiavacchi_akhavan_1992}; \cite{Coleman1996,Scotti2001,he_seddighi_2015,weng_boij_hanifi_2016,Jung2017,Sundstrom2018self,lozano2020non} and more detailed background is provided in \S \ref{sec:stokeslayer}. These studies, in which a laminar Stokes layer  is observed near the wall, motivate the non-equilibrium model introduced in \S \ref{sec:stokeslayer}. These studies also discuss how turbulence structure itself is affected by the unsteadiness above the viscous sublayer, however, we leave  modeling of such effects for future work. 

After introducing the two basic ingredients for the new wall model, in \S \ref{sec:implementations} we address practical implementation issues. 
Applications to equilibrium and non-equilibrium channel flows are described in \S \ref{sec:channel}. Various features of the model are presented via time series at individual points as well as contour plots of modeled wall stress components. The tests in non-equilibrium  flow are performed for channel flow upon which a very strong sudden spanwise pressure gradient (SSPG) is  applied.  Comparisons with DNS results by \cite{lozano2020non} are also included.  Summary and conclusions are presented in \S \ref{sec:conclusions}. 

\section{Lagrangian relaxation towards equilibrium (LaRTE) wall model}
\label{sec:LaRTEmodel}

\begin{figure}
\centering
\includegraphics[width=0.8\linewidth]{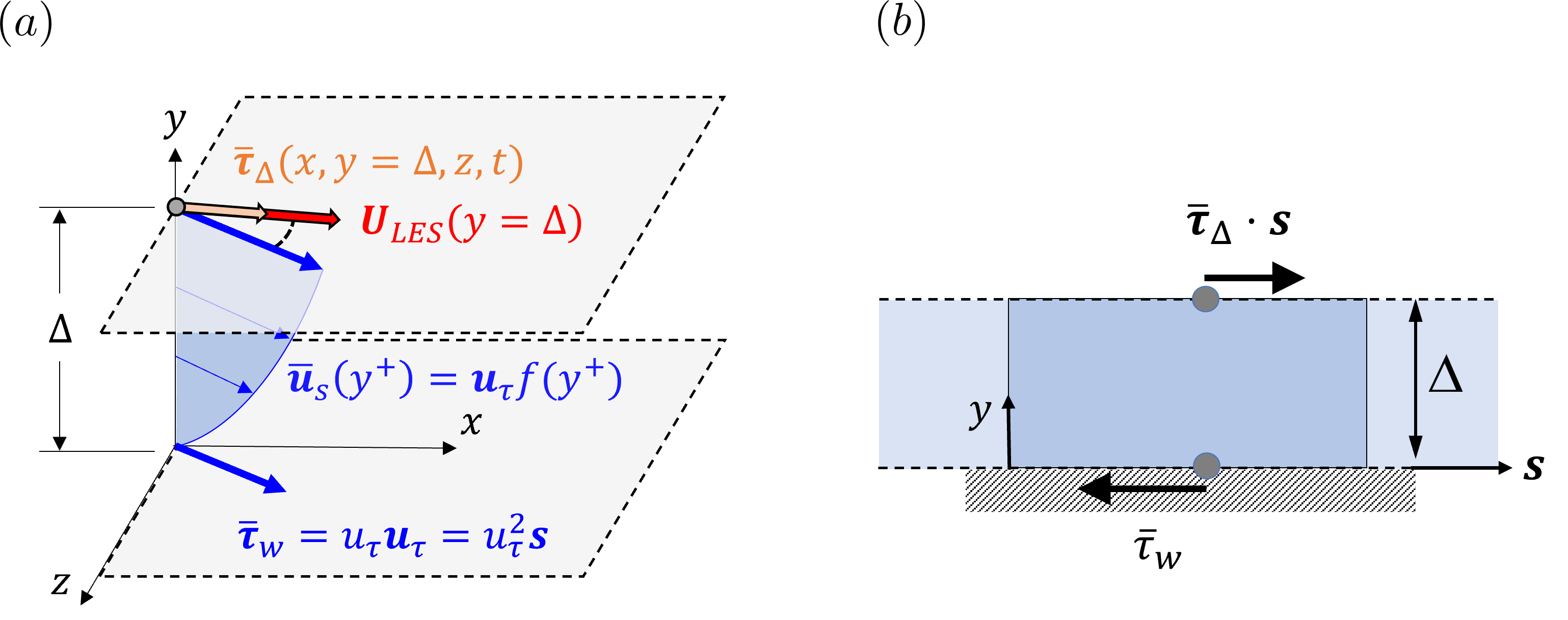}
\caption{(a) Sketch of assumed inner velocity profile (in blue) representing a quasi-equilibrium RANS solution in the inner layer, responding to an outer ``applied'' total shear stress $\overline{\bm{\tau}}_\Delta$ at the wall model height at $y=\Delta$. (b) Sketch of stresses acting on the fluid layer between $y=0$ and $y=\Delta$, leading to inertia term with response time-scale $T_s$ proportional to $\Delta$.}
\label{fig:sketch}
\end{figure}

Following the ideas underlying the integral wall model \citep{yang2015integral} we assume that between the wall and the wall model height (an LES grid point at a distance $\Delta$ away from the wall there exists a quasi-equilibrium mean velocity profile (see figure \ref{fig:sketch}(a)). An overline denotes the corresponding averaging operation, which may be interpreted as a horizontally grid filtered quantity at the LES scale in the wall-parallel plane, and additional (implicit) temporal averaging whose properties will become apparent from the derivation itself. The key assumption underlying the proposed wall model is that in the horizontal (wall parallel, $x-z$) plane, the  mean velocity $\overline{\bm u}_s = \overline{u} \, \ihat + \overline{w} \, \khat$  ($\ihat$ and $\khat$ are the two unit vectors on the wall) can be written according to
\be
\overline{\bm u}_s(x,y,z,t)  =  {\bm u}_{\tau}(x,z,t) \,\ f(y^+),
\label{eq:mainansatz}
\ee
where ${\bm u}_\tau(x,z,t)$ is the friction velocity vector and is a slowly varying function of the horizontal positions $x,z$ and time $t$. The characteristic time-scale characterizing what is termed ``slow'' evolution is not prescribed a-priori but will be shown to arise directly from the assumption of quasi-equilibrium. The inner similarity function $f(y^+)$ with $y^+ = y u_\tau/\nu$ is the assumed velocity profile in inner units, and $u_{\tau} = |{\bm u}_{\tau}|$. Typically $f(y^+)$  includes a linear region near the wall merging with a logarithmic portion above the buffer layer but the precise shape of $f(y^+)$ is not important at initial stages of development. We remark that in the present work we deal exclusively with smooth planar walls.  

The full quasi-steady velocity is then given by 
\be \label{eq:vel_decomp}
\overline{\bm u} = \overline{\bm u}_s + \overline{v} \jhat, 
\ee
where $\overline{v}$ is the wall-normal velocity and $\jhat$ the unit vector in the $y$-direction. 
The friction velocity vector ${\bm u}_{\tau} = u_{\tau x}{\ihat} + u_{\tau z}{\khat}$ is defined such that the (kinematic) wall stress vector $\overline{\bm\tau}_w$ (its two components in the wall plane) is given by
\be
\overline{\bm\tau}_w = u_{\tau} {\bm u}_{\tau},
\ee
i.e. $\overline{\bm u}_s$ and ${\bm u}_{\tau}$ are in the same direction as $\overline{\bm\tau}_w$. This direction will be represented by unit vector ${\bm s}$ (that also can depend on $x,z,t$), i.e. ${\bm u}_\tau = u_\tau {\bm s}$ and $\overline{\bm u}_s = u_\tau\, f(y^+) \, {\bm s}$
(figure \ref{fig:sketch}(a)). 

Next, we aim to derive an evolution equation for the friction velocity vector ${\bm u}_{\tau}(x,z,t)$ that is consistent with the RANS evolution for $\overline{\bm u}$:    
\begin{equation}
\frac{\partial \overline{\bm{u}}}{\partial t} + \bm{\nabla} \cdot \left( \overline{\bm{u}}   \, \overline{\bm{u}} \right)
= - \frac{1}{\rho} \bm{\nabla}\overline{p} 
+  \bm{\nabla} \cdot \left[(\nu+\nu_T)  (\bm{\nabla} \overline{\bm{u}}+\bm{\nabla} \overline{\bm{u}}^\top) \right],
\label{eq:urans}
\end{equation}
where $\nu_T(x,y,z,t)$ is the position-dependent eddy viscosity associated with the RANS model being considered and $\overline{p}(x,z,t)$ is the quasi-equilibrium pressure with no wall-normal dependence, consistent with boundary layer approximations. The momentum equation for the  wall-parallel velocity $\overline{\bm{u}}_s$ (2 components) reads
\begin{align}
\frac{\partial \overline{\bm{u}}_s}{\partial t} + \bm{\nabla}_h \cdot \left( \overline{\bm{u}}_s   \, \overline{\bm{u}}_s \right) + \partial_y ( \overline{v} \, \overline{\bm{u}}_s ) 
= - \frac{1}{\rho} \bm{\nabla}_h\overline{p} 
&+ \frac{\partial}{\partial y} \left[ (\nu+\nu_T) \frac{\partial \overline{\bm{u}}_s}{\partial y} \right] \nonumber \\
&+ \bm{\nabla}_h \cdot \left[(\nu+\nu_T) \left(  \bm{\nabla}_h \overline{\bm{u}}_s+\bm{\nabla}_h \overline{\bm{u}}_s^\top \right)  \right],
\label{eq:horizmom}
\end{align}
where $\bm{\nabla}_h=\partial_x \ihat +\partial_z \khat$ represents the horizontal gradients on the $x-z$ wall plane, and diffusion cross terms involving the (small) vertical velocity $\overline{v}$ have been neglected. 
Into this equation we replace the main ansatz (\ref{eq:mainansatz}). And, following the logic by \cite{chung2009large} and the integral wall model by \cite{yang2015integral}, we integrate from $y=0$ to the wall-model height at $y=\Delta$. The procedure will be illustrated via the first term, the Eulerian time derivative $\partial_t \overline{\bm u}_s$,  but similar steps can be applied to the advective derivative terms also invoking the continuity equation as shown in Appendix \ref{app:advection}. The key steps for the time derivative read as follows: 
\begin{align}
\frac{\partial}{\partial t} \int_0^\Delta \overline{\bm{u}}_s dy 
&= \frac{\partial}{\partial t} 
\left[\, {\bm s} \, \int_0^\Delta u_\tau \, f\left(\frac{y u_\tau}{\nu} \right) \, dy   \right] \nonumber \\
&= {\bm s} \int_0^\Delta \frac{\partial u_\tau}{\partial t} \left( f(y^+) + u_\tau f^\prime(y^+) \frac{y}{\nu} \right) dy  + \frac{\partial {\bm s}}{\partial t} \int_0^\Delta u_\tau \, f(y^+) \, dy.
\end{align}
As recognized by \cite{chung2009large} in their derivation of an integral boundary layer equation-based wall model, the first integral on the RHS can be rewritten with $d/dy^+[y^+ f(y^+)]$ as integrand, resulting in
\begin{align}
\frac{\partial}{\partial t} \int_0^\Delta \overline{\bm{u}}_s dy 
&= {\bm s} \frac{\partial u_\tau}{\partial t} \int_0^{\Delta^+} \frac{d}{dy^+} \left[ y^+f(y^+) \right] dy^+ \frac{\nu}{u_\tau}  + u_\tau \,  \frac{\partial {\bm s}}{\partial t} \int_0^\Delta f(y^+) \, dy \nonumber \\
&= {\bm s} \frac{\partial u_\tau}{\partial t} \Delta f(\Delta^+) + u_\tau \,  \frac{\partial {\bm s}}{\partial t} \int_0^\Delta f(y^+) \, dy  \nonumber \\
&=  \frac{\partial (u_\tau{\bm s}) }{\partial t} \, \Delta f(\Delta^+) + u_\tau \,  \frac{\partial {\bm s}}{\partial t} \, \left(\int_0^\Delta [f(y^+)-f(\Delta^+)] \, dy \right).
\end{align}
The last term motivates definition of a ``cell displacement thickness''
\be \label{delta_star_Delta_def}
\delta^*_\Delta = \int_0^\Delta \left(1-\frac{\overline{u}_s(y)}{\overline{u}_s(\Delta)} \right) \, dy \,\,\,\to  \,\,\,
\frac{\delta^*_\Delta}{\Delta} = 
\frac{1}{\Delta^+} \, \int_0^{\Delta^+} \left(1-\frac{f(y^+)}{f(\Delta^+)} \right) \, dy^+,
\ee
analogous to the boundary-layer displacement thickness but integrated only up to $y=\Delta$. Finally, the Eulerian time derivative term can be written according to
\be 
\frac{\partial}{\partial t} \int_0^\Delta \overline{\bm{u}}_s dy =  \Delta f(\Delta^+) \, \frac{\partial {\bm u}_\tau}{\partial t} - u_\tau f(\Delta^+) \, \delta^*_\Delta \, \frac{\partial {\bm s}}{\partial t}.
\label{eq:dLdt}
\ee
Integration of the advective term, i.e. $\int_0^\Delta  \bm{\nabla}_h \cdot \left( \overline{\bm{u}}_s \, \overline{\bm{u}}_s \right) + \partial_y ( \overline{v} \, \overline{\bm{u}}_s ) \, dy$ requires the mean vertical velocity at $y=\Delta$, since $\int_0^\Delta  \partial_y ( \overline{v} \, \overline{\bm{u}}_s ) \, dy = \overline{v}(\Delta) \overline{\bm{u}}_s(\Delta) {\bm s}$ (and ${\bm s}$ does not depend on $y$). Using the continuity equation $\partial_s \overline{u}_s + \partial_y \overline{v}=0$ we obtain 
\begin{align}
\overline{v}(\Delta) = - \int_0^\Delta \partial_s \overline{u}_s(y) \, dy 
&= - \frac{\partial}{\partial s} \left[  u_\tau \int_0^\Delta f(y^+) \,dy \, \right] \nonumber \\ 
&= - \frac{\partial u_\tau}{\partial s}  \int_0^\Delta \left[ f(y^+) + y^+ f'(y^+) \right] dy   = - \frac{\partial u_\tau}{\partial s} \Delta f(\Delta^+).
\label{eq:vertvel}
\end{align}
As further shown in detail in Appendix \ref{app:advection} the entire integral of the advective term can then be written as
\be
\int_0^\Delta \left[\bm{\nabla}_h \cdot \left( \overline{\bm{u}}_s \, \overline{\bm{u}}_s \right) + \partial_y ( \overline{v} \, \overline{\bm{u}}_s ) \right] \, dy \, = \, \Delta \, f(\Delta^+) \, {\bm V}_\tau \cdot \bm{\nabla}_h {\bm u}_\tau,  
\ee
where
\be 
 {\bm V}_\tau = \left(1-\frac{\delta^*_\Delta}{\Delta} - \frac{\theta_\Delta}{\Delta}\right) \, f(\Delta^+) \,  {\bm u}_\tau
 \label{eq:defVtau}
\ee
is the advective velocity and a ``cell momentum thickness'' 
\be \label{theta_Delta_def}
\theta_\Delta =  \int_0^\Delta \frac{\overline{u}_s(y)}{\overline{u}_s(\Delta)} \left(1-\frac{\overline{u}_s(y)}{\overline{u}_s(\Delta)} \right) \, dy  \,\,\,\to  \,\,\, \frac{\theta_\Delta}{\Delta} =  \frac{1}{\Delta^+} \, \int_0^{\Delta^+} \frac{f(y^+)}{f(\Delta^+)} \left(1-\frac{f(y^+)}{f(\Delta^+)} \right) \, dy^+,
\ee
has been introduced arising from the integrals of quadratic advection terms. 

Integrating each term in equation (\ref{eq:horizmom}) between $y=0$ and $y=\Delta$, it is convenient to define the total (molecular viscous + turbulent viscous) shear stress at $y=\Delta$ according to $\overline{\bm\tau}_\Delta = (\nu+\nu_T)\partial \overline{\bm u}_s/\partial y$. Collecting terms, replacing ${\bm u}_\tau = u_\tau {\bm s}$, and dividing the entire equation by $\Delta f(\Delta^+)$, the evolution equation for the friction velocity vector can now be written according to
\be 
\frac{\partial  {\bm{u}}_\tau}{\partial t} + {\bm V}_\tau \cdot \bm{\nabla}_h {\bm{u}}_\tau 
=  \frac{1}{T_s} \left[ \frac{1}{u_\tau}\left(-\frac{\Delta}{\rho} \bm{\nabla}_h \overline{p} + \overline{\vct{\tau}}_\Delta \right) - {\bm u}_\tau\right]\, + \, u_\tau \frac{\delta^*_\Delta}{\Delta} \frac{\partial {\bm s}}{\partial t} \,+\,\frac{1}{\Delta f(\Delta^+)} \,\, {\bm \nabla}_h \cdot {\cal \bm D}_\tau,
\label{eq:Lagrelax} 
\ee
where  $T_s$ is given by
\be \label{Ts}
T_s = f(\Delta^+) \, \frac{\Delta }{u_\tau}.
\ee
It represents a time-scale that arises from the derivation of equation (\ref{eq:Lagrelax}) and does not require additional ad-hoc assumptions.

The horizontal diffusion flux tensor integrated in the vertical direction, ${\cal \bm D}_\tau$,  is  defined according to 
\be {\cal \bm D}_\tau = \int_0^\Delta (\nu+\nu_T) \left(  \bm{\nabla}_h \overline{\bm{u}}_s+\bm{\nabla}_h \overline{\bm{u}}_s^\top \right)    \,  dy,
\ee
and is further detailed in Appendix \ref{app:diffusion}. When coupled with appropriate models for $\overline{\vct{\tau}}_\Delta$, $f(\Delta^+)$, $\delta^*_\Delta$, $\theta_\Delta$, and ${\cal \bm D}_\tau$, we refer to  equation (\ref{eq:Lagrelax}) as the evolution equation underlying the LaRTE wall model.

\subsection{Discussion}
\label{sec:discussion}

For the sake of initial discussion, it is instructive to consider a simplified form for the (e.g.) streamwise 
$x$-component of equation (\ref{eq:Lagrelax}) and for now neglecting the pressure gradient and diffusive terms, as well as the direction change term $\partial{\bm s}/\partial t$. Under these simplifying conditions, equation (\ref{eq:Lagrelax}) can be written as
\be
\frac{ {d}_s u_{\tau x}}{dt \,\,}  = \frac{1}{T_s}\left(  \overline{ {\tau}}_{\Delta x}/u_\tau - {u}_{\tau x} \right),
\label{relax1_ux}
\ee
with ${d}_s/dt=\partial_t + {\bm V}_\tau \cdot {\bm \nabla}_h$ representing a Lagrangian time derivative on the surface. In this form, it becomes apparent that the model represents a Lagrangian relaxation dynamics, with $T_s$ serving as the relaxation time-scale for how the friction velocity component ${u}_{\tau x}$  approaches the stress at the wall-model grid point in LES ($\overline{{\tau}}_{\Delta x}$) (the latter divided by the friction velocity magnitude $u_\tau$).  For the present discussion, neglecting $\overline{\tau}_{w z}$, i.e.~with $\overline{\tau}_{wx}=u_\tau u_{\tau x} =  u_{\tau x}^2$, we see by multiplying (\ref{relax1_ux}) by $u_{\tau x}$ that in terms of the wall stress the Lagrangian relaxation equation can equivalently be written as
\be
\frac{d_s \overline{\tau}_{w x}}{dt} = \frac{2}{T_s} \left( \overline{\tau}_{\Delta x} - \overline{\tau}_{w x} \right),
\label{eq:relax1_tauwx}
\ee
showing that for the stress the relaxation time-scale is $T_s/2$. This time scale was originally derived by \cite{chung2009large} where they also showed the wall stress tends (in an Eulerian sense) towards its steady state value at a rate corresponding to this time scale (in their work $1/(\Lambda \widetilde{\eta}_0)$ is equivalent to $T_s/2$ shown here).

It can be seen from equation (\ref{eq:dLdt}) that the relaxation time scale $T_s$ arises from integrating the assumed velocity profile between $y=0$ and $y=\Delta$ \citep[this integral is similar to the term $\partial L/\partial t$ that arises in the iWMLES approach by][]{yang2015integral}.  As a result of the analysis, $T_s$ is proportional to the volume per unit area ($\Delta$) in the fluid layer under consideration, between the wall and the  wall-model height $y=\Delta$. It represents the inertia of the fluid in that layer and can be seen to cause a time-delay between the  stress at $y=\Delta$ and at the wall, under unsteady conditions. The thicker the layer (large $\Delta$), the more the time delay due to added fluid inertia. Conversely, the stronger the turbulence (large $u_\tau$), the faster the relaxation leading to a smaller time-delay.  We note that in  \cite{yang2015integral} for the iWMLES approach, an explicit time-filtering at a time-scale $\sim \Delta/\kappa u_\tau$ was introduced, operationally similar but shorter than $T_s$ by a factor $\kappa f(\Delta^+) = \kappa \overline{u}(\Delta)/u_\tau$. Here, such temporal relaxation behavior has been derived formally from the momentum equation (unsteady RANS) and the assumed validity of a quasi-equilibrium velocity profile (Eq. \ref{eq:mainansatz}).  

Also, for the full vector problem with two stress components, we remark that when attempting to write the relaxation equation in terms of wall-parallel components $\overline{\tau}_{wx}$ and $\overline{\tau}_{wz}$ instead of  friction velocity vector components $u_{\tau x}$ and $u_{\tau z}$ (or ${\bm u}_{\tau}$), the resulting equation is far less intuitive compared to the relatively simple form of equation (\ref{eq:Lagrelax}). The latter resembles a standard transported vector field equation with a relaxation source term and a diffusion term (only the $\partial {\bm s}/\partial t$ term is non-standard) and is therefore much preferable. 

Finally, we note that when including the pressure gradient term $\nabla_h \overline{p}$, equation (\ref{eq:Lagrelax}) shows that the implied relaxation dynamics  is for how ${\bm u}_\tau$ approaches the total cell forcing (vector) term $ (-\rho^{-1}{\bm \nabla}_s \overline{p}\, \,\Delta + \overline{{\bm \tau}}_\Delta)/u_\tau$.

\subsection{Closure for the total stress at the wall-model height}
\label{sec:taudeltaclosure}
The relaxation wall model requires specification of the total (molecular viscous + turbulent) stress $\overline{\vct{\tau}}_{\Delta}$ at $y=\Delta$ as function of known LES quantities there, such as the LES velocity ${\bm U}_{\rm LES}$ at the wall-model point. I.e. we denote $\tilde{\bm u}(y=\Delta) = {\bm U}_{\rm LES}$, where $\tilde{\bm u}({\bm x},t)$ is the velocity field being solved in LES.  We argue that introducing a closure model for the total stress there ($\overline{\vct{\tau}}_{\Delta}$) is more appropriate than closing the stress at the wall, since the wall is at a different position than the position at which the LES velocity is known. For now the simplest approach, which we shall adopt in this paper, is to use a standard ``equilibrium'' RANS closure to relate the known velocity ${\bm U}_{\rm LES}(y=\Delta)$ to the total (viscous plus turbulent) shear stress $\overline{\vct{\tau}}_{\Delta}$ at the same position. In this way, we can connect the new LaRTE wall model to the traditional equilibrium wall model: the latter is obtained simply by letting $T_s \to 0$, in which case the wall stress is set equal to the total stress at $y=\Delta$. This equality of stresses would be justified under full equilibrium conditions, i.e. if the turbulence small-scale unresolved motions operated on much shorter time-scales than the macroscopic variables (time-scale separation) and $T_s \to 0$ would be appropriate. However, for turbulence that lacks scale separation under quasi-equilibrium conditions with temporal and spatial variations and pressure gradient effects, the assumption that $T_s \to 0$ and equality of stresses are not formally justified. Then the relaxation equation can be solved instead.   

Next, we describe the proposed closure for the stress $\overline{\bm{\tau}}_\Delta$.  We use the approach developed in \cite{meneveau2020note} in which an equilibrium layer partial differential equation is numerically integrated and various fitting functions are developed for this solution. The model of \cite{meneveau2020note} is expressed in the form of a friction Reynolds number that depends on a $U_{\rm LES}$-based Reynolds number and a dimensionless pressure gradient parameter via a dimensionless fitting function:
\be \label{retd_fit}
\frac{\langle u_\tau \rangle \Delta }{\nu} \, = \, Re_{\tau\Delta}^{\rm pres}(Re_\Delta, \psi_p), \,\,\, {\rm where} \,\,\,\, 
Re_\Delta = \frac{U_{\rm LES} \Delta}{\nu} \,\,\, {\rm and} \,\,\, \psi_p = \frac{1}{\rho} (\nabla_h P \cdot \hat{\bm e}_u) \, \frac{\Delta^3}{\nu^2}
\ee
where the superscript ``pres'' indicates that the fit includes pressure gradient dependence (unlike $Re_{\tau \Delta}^{\rm fit}(Re_\Delta)$ also provided in \cite{meneveau2020note}). This fit is repeated in Appendix C for completeness. Also, $\hat{\bm e}_u={\bm U}_{\rm LES}/|{\bm U}_{\rm LES}|$ is the unit vector in the ${\bm U}_{\rm LES}$ direction. The pressure gradient $\nabla_h P$ represents a steady or very low-frequency background pressure gradient, included in the full equilibrium part of the dynamics. 
Details of how  $\nabla_h P$ is determined in simulations are provided later, in \S \ref{sec:pg_decomposition}. 
The fit for $Re_{\tau\Delta}^{\rm pres}(Re_\Delta,\psi_p)$ provided in \cite{meneveau2020note} was obtained by numerically integrating the simple steady 1D RANS equations that, unlike Eq. \ref{eq:urans} did not include time-dependence and can thus be characterized as a ``fully equilibrium model'' (as opposed to quasi-equilibrium) assumption. 
The friction velocity $\langle u_\tau \rangle$ is the corresponding ``full equilibrium'' friction velocity.  Under these conditions, the full equilibrium vertically integrated momentum equation implies that $0 = -\Delta \,\nabla_h P/\rho + \overline{\tau}_\Delta - \langle u_\tau \rangle^2$, i.e. $\langle u_\tau \rangle^2$ obtained from applying  Eq. \ref{retd_fit} represents a model for the combined $\overline{\tau}_\Delta - \Delta \, \nabla_h \, P/\rho$. 

Following usual practice of equilibrium wall models, we assume that the total stress  modeled by the fitted equilibrium expression is aligned with the LES velocity at the first grid point and write
\be
\overline{\bm{\tau}}_\Delta - \frac{1}{\rho} \Delta {\bm \nabla}_h P =  \frac{1}{2} \, c_{f}^{\rm wm} \, U_{\rm LES}^2 \,\, \hat{\bm e}_u =  \left( Re_{\tau\Delta}^{\rm pres} \, \, \frac{\nu}{\Delta} \right)^2 \, \hat{\bm e}_u 
\label{eq:modelfittau}
\ee
where 
$Re_{\tau\Delta}^{\rm pres}=Re_{\tau\Delta}^{\rm pres}(Re_\Delta, \psi_p)$ is the fit. Note that the latter is related to the ``equilibrium wall model friction factor'' $c_{f}^{\rm wm}$  according to $c_{f}^{\rm wm} = 2 \left( {Re_{\tau\Delta}^{\rm pres}}/{Re_\Delta}\right)^2$. 

Finally then, the two-dimensional PDE governing the evolution of the friction velocity vector in the LaRTE model reads as follows:
\be 
\frac{\partial  {\bm{u}}_\tau}{\partial t} + {\bm V}_\tau \cdot \bm{\nabla}_h {\bm{u}}_\tau 
=  \frac{1}{T_s} \left[ \frac{1}{u_\tau}\left(-\frac{\Delta}{\rho} \bm{\nabla}_h \overline{p}' + (Re_{\tau \Delta}^{\rm pres} \, \nu/\Delta)^2 \hat{\bm e}_u \right) - {\bm u}_\tau\right]\, + \, u_\tau \frac{\delta^*_\Delta}{\Delta} \frac{\partial {\bm s}}{\partial t},
\label{eq:Lagrelax2} 
\ee
where $\overline{p}'$ is the pressure fluctuation that excludes the very slow background pressure gradient based on $P$ and is further described in \S \ref{sec:pg_decomposition}. Also note that we have neglected the horizontal diffusion term in writing this equation (further discussion of horizontal diffusion is provided in \S \ref{sec:Lagtimederivativenumerics}). The solution to Eq. \ref{eq:Lagrelax2} is then used to determine the quasi-equilibrium part of the wall stress $\overline{\bm{\tau}}_w=u_\tau {\bm u}_\tau$. 

In practice for flows such as channel flow or zero pressure gradient boundary layers that do not display major non-equilibrium effects, one would not expect to see noticeably different results in overall flow statistics, whether one applies the equilibrium wall model instantaneously at the wall as is usually done, or if one applies the proposed Lagrangian time relaxation, i.e. with some time delay and smoothing. However, the new formulation enables us to operationally separate the model self-consistently into a part that genuinely represents quasi-equilibrium dynamics, and a remainder for which additional modeling is needed. As introduced in \S \ref{sec:stokeslayer}, the total modeled wall stress may also include a non-equilibrium component $\bm{\tau}_w^{\prime \prime}$, representing additional   contributions to the wall stress that do not arise from the quasi-equilibrium dynamics encapsulated in Eq.  \ref{eq:Lagrelax} and \ref{eq:Lagrelax2}.  For example, in \S \ref{sec:stokeslayer} we introduce an additional term $\bm{\tau}_w^{\prime \prime}$   intended to capture the non-equilibrium laminar response to a rapidly changing pressure gradient in the viscous sublayer. But first, in the next section we present an a-priori data analysis  to study properties of several averaging time-scales.

\subsection{A-priori analysis of quasi-equilibrium dynamics in channel flow}
\label{sec:apriori}
Here we examine channel flow DNS data at $Re_\tau = $1{,}000 to show that the time-scale $T_s$ identified in Eq. \ref{Ts} provides a self-consistent decomposition of the flow into quasi-equilibrium and non-equilibrium (the remainder) components. DNS data were obtained from the Johns Hopkins Turbulence Database \citep{JHTDBwebsite} for the channel flow $Re_\tau=$1{,}000 dataset \citep{graham2016web}. The velocity data was Gaussian filtered horizontally at scale $\Delta_x^+=\Delta_z^+=196$, commensurate to the LES grid resolution for simulations considered later in this paper. The velocity was collected for all times available, $0\leq t u_\tau / h \leq 0.3245$, at a single point, $(x_0,z_0)$, over a wall-normal height $0\leq y^+ \leq \Delta^+ \approx 34$, close to the height of the first LES grid point that will serve as the wall-model height.

\begin{figure}
\centering
\includegraphics[width=0.8\linewidth]{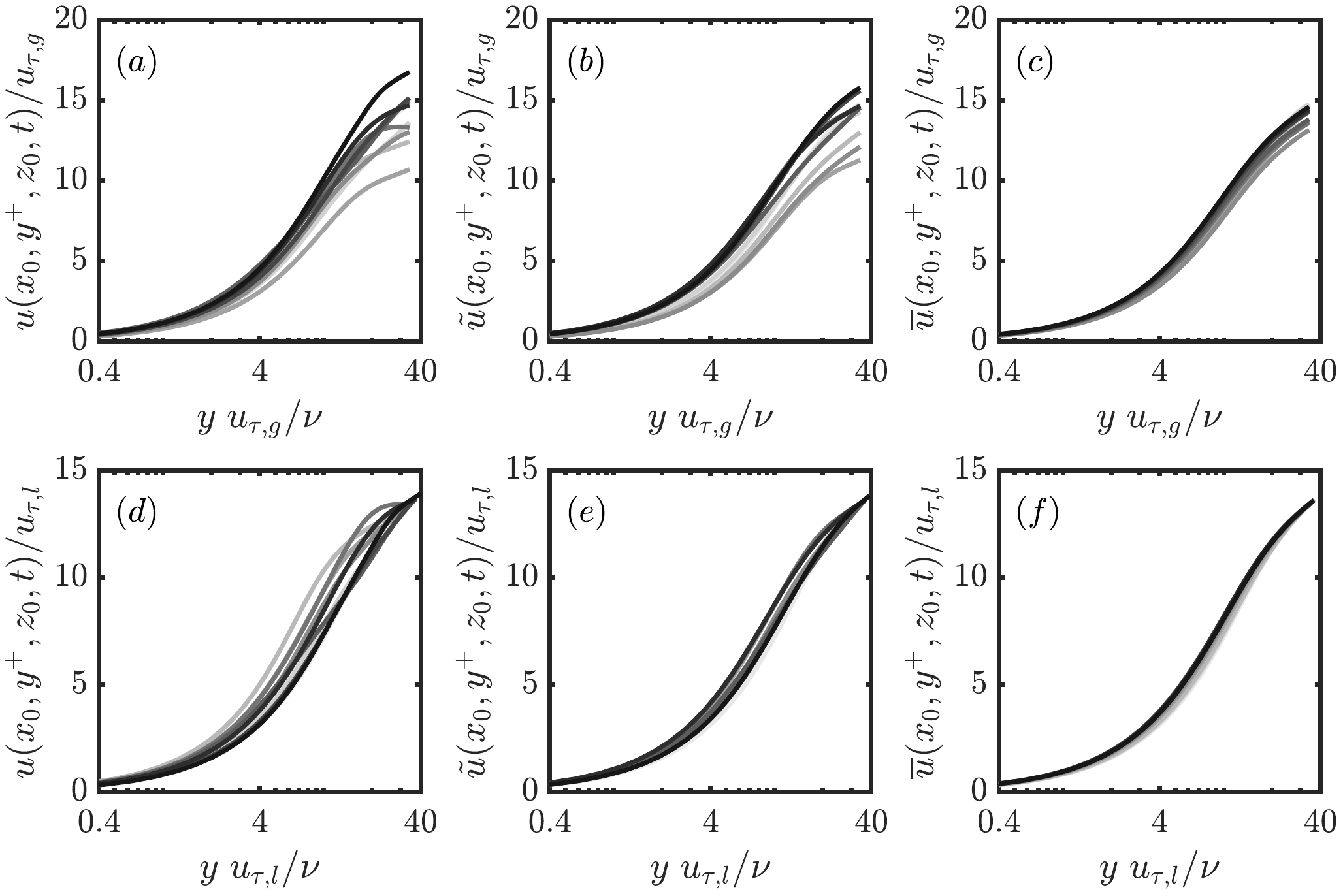}
\caption{Velocity profiles at a single point $(x_0,z_0)$ at various times, from a-priori tests from  DNS data, Gaussian filtered in the horizontal directions at $\Delta_x^+=\Delta_z^+=196$. Different lines represent different times, lighter line color corresponds to earlier time, separated by $t u_{\tau,g}/h=0.13$. Profiles are normalized using the global friction velocity $u_{\tau,g}$ in $(a-c)$ while $(d-f)$ use the local friction velocity $u_{\tau,l}$ as averaged over the same time filtering scale as used for the profile. $(a,d)$: no time filter; $(b,e)$: exponentially time filtered with $T_1=\Delta/u_{\tau,g}$ time filter;  $(c,f)$: exponentially time filtered using $T_s = \Delta f(\Delta u_{\tau,g}/\nu)/u_{\tau,g}$ consistent with the LaRTE approach. The $y^+$ dependence in the vertical axis matches that of the horizontal axis (i.e. $y^+ = y u_{\tau,g}/\nu$ for $(a-c)$ and $y^+ = y u_{\tau,l}/\nu$ for $(d-f)$).}
\label{fig:dns_apriori}
\end{figure}

The DNS velocity is then temporally filtered below $y=\Delta$ using a one-sided exponential time filter, computed as $\tilde{u}^n = \epsilon u^n + (1-\epsilon)\tilde{u}^{n-1}$, where  $\epsilon = \delta t/T$, $\delta t$ is the time-step size, $n$ is the time step index, and $T$ is the averaging time scale. Three different averaging time scales are considered to see which time scale is most consistent with quasi-equilibrium assumptions. Quasi-equilibrium is satisfied when the filtered velocity profile collapses to $\widetilde{u} = u_{\tau,l} f(y u_{\tau,l}/\nu)$ for all time, where $u_{\tau,l}(x,z,t)$ is the local friction velocity. The local friction velocity is computed using $u_{\tau,l} = \nu/\Delta \, Re_{\tau}^{fit}(U_{LES}\Delta/\nu)$ where $Re_{\tau}^{fit}$ is the inverted law of the wall fit from \cite{meneveau2020note} and $U_{LES} = \widetilde{u}(x,y=\Delta,z,t)$ is the filtered velocity at the wall-model height. In the limit $T \to \infty$, the velocity profile is static and thus full equilibrium is achieved. In this limit the local friction velocity tends towards the global friction velocity, computed as $u_{\tau,g} = \sqrt{(h/\rho)(-d\langle p \rangle/dx)}$ where $d\langle p \rangle/dx$ is the bulk pressure gradient forcing for channel flow. Inner units normalization of the velocity profile using $u_{\tau,g}$ is shown in the top row of figure \ref{fig:dns_apriori} whereas normalization with $u_{\tau,l}$ is shown in the bottom row of figure \ref{fig:dns_apriori}. In order, from left to right in figure \ref{fig:dns_apriori}, the averaging time scales considered are: panels $(a,d)$:
no temporal filtering, i.e. entirely local; panels $(b,e)$: intermediate time scale $T_1=\Delta/u_{\tau,g}$; and panels  $(c,f)$: LaRTE predicted time-scale $T_s = \Delta f(\Delta u_{\tau,g}/\nu)/u_{\tau,g}$. Note that $T_1$ is similar to what is used in \cite{yang2015integral}, however off by a factor $\theta/\kappa$ with $\theta=1$ used for their work. They also mentioned that a longer time scale may be needed, even suggesting a minimum of $\theta = 5$ which yields a coincidentally rather similar time scale as $T_s$ for $\Delta^+ = 34$. We stress that here $T_s$ is derived based on a momentum balance and  does not require tunable parameters as was the case in \cite{yang2015integral}. 

From figure \ref{fig:dns_apriori}($f$) it is clear that filtering with the relaxation time scale $T_s$ most closely satisfies quasi-equilibrium assumptions, as it almost completely collapses to the law of the wall when normalized with the local friction velocity. This a-priori test therefore provides justification that the relaxation towards equilibrium wall model, which responds within the relaxation time scale $T_s$ as discussed in \S \ref{sec:discussion}, is consistent with quasi-equilibrium assumptions.

\section{Non-equilibrium laminar fast response model}  
\label{sec:stokeslayer}
So far we have developed a model for $\overline{\bm{\tau}}_w$, the quasi-equilibrium part of the wall stress, which responds to external conditions (changes in velocity and pressure gradients at the wall-model height) at a time-scale $T_s$ consistent with the assumption of quasi-equilibrium. In order to supplement the quasi-equilibrium Lagrangian relaxation model with additional physics, we now focus on the rapid response of the inner-most part of the boundary layer, the viscous sublayer. The response of near-wall structures of turbulence to rapidly changing pressure gradients has been studied extensively in the past.
\cite{jung_mangiavacchi_akhavan_1992}, \cite{Karniadakis2003}, \cite{Quadrio2003}, \cite{Ricco2012}, and \cite{Yao2019} studied spanwise wall-oscillations due to its drag-reducing capabilities. \cite{Vardy2003transient} and \cite{Vardy2015applicability} attempt to understand the wall shear stress for water hammer pipe flows. Experimental and numerical studies of pulsatile flows have been performed by \cite{Scotti2001}, \cite{Tardu2005experiments}, \cite{tardu2010wall}, \cite{weng_boij_hanifi_2016}, \cite{Sundstrom2018characteristics}, and \cite{Cheng2020forcing}. Streamwise accelerating flows were considered by \cite{he_jackson_2000}, \cite{greenblatt_moss_2004}, \cite{He2008computational}, \cite{He2011accelerating}, \cite{He2011early}, \cite{he_seddighi_2013}, \cite{he_seddighi_2015}, \cite{Jung2012}, \cite{Jung2017}, \cite{Sundstrom2017response}, and \cite{Sundstrom2018self}. Then there are studies with step changes in either the wall boundary condition or in the pressure forcing such as the sudden spanwise wall movement of \cite{Coleman1996}, \cite{tang_akhavan_2016}, and \cite{abe_2020}, the sudden spanwise pressure gradient of \cite{Moin1990direct} and \cite{lozano2020non}, or a change in the direction of the pressure gradient forcing (for a variety of directions) as done in \cite{wiart2018}. One of the common observations amongst all of these flows is the existence of a laminar Stokes layer near the wall. For pulsatile flows, the wall stress deviation from its steady state value follows the solution to Stokes's second problem exactly for high frequency oscillations (\cite{weng_boij_hanifi_2016}). For streamwise accelerating flows, the wall stress deviation from its initial value follows the solution to Stokes's first problem during the first stage of the acceleration (\cite{he_seddighi_2015}, \cite{Jung2017}, \cite{Sundstrom2018self}). \cite{Sundstrom2018similarity} even showed that the wall stress for low frequency pulsations follows the Stokes solution during the acceleration phase of the pulse. They further showed the similarity of the wall stress during the acceleration phase of pulsatile flows with the initial phase of streamwise accelerating flows. For the sudden spanwise wall movement and sudden spanwise pressure gradient, the spanwise velocity and wall stress follows Stokes's first problem during the early response \citep{Coleman1996,abe_2020,lozano2020non}. 
 
We here use concepts inspired by these prior works to complement the quasi-equilibrium model presented in \S \ref{sec:LaRTEmodel}. We decompose the velocity $\tilde{\bm{u}}$ (this velocity is spatially filtered in the 2D horizontal plane but not time filtered, and so it may still contain strong time and $y$-dependent deviations from the quasi-equilibrium profile $\overline{\bm{u}}$)  according to $\tilde{\bm{u}} = \overline{\bm{u}} + \tilde{\bm{u}}''$ where $\tilde{\bm{u}}''$ is the non-equilibrium velocity to be modeled here. The deviations from the quasi-equilibrium velocity distribution $\overline{\bm{u}}$, such as the deviations visible in figures \ref{fig:dns_apriori}(a) and (d), can arise from a variety of sources such as time-dependent turbulent fluctuations and pressure gradients. The fastest changing pressure gradient fluctuations can induce oscillatory flow conditions even in the viscous sublayer, not unlike those involved in the Stokes first and second problems. Since for the quasi-laminar part of the flow in the viscous sublayer an analytical solution can be developed, we aim now to model that part of the non-equilibium wall stress arising directly from the response of the laminar sublayer to rapid pressure gradient fluctuations. We denote the corresponding laminar velocity response as $\tilde{\bm{u}}_{l}''$ 
where subscript ``$l$'' stands for laminar component.

In the viscous sublayer the linear terms of the Navier-Stokes equation dominate and hence we argue that
 $\tilde{\bm{u}}_{l}''$ obeys
\be
\frac{\partial \tilde{\bm{u}}_l''}{\partial t} = - \frac{1}{\rho}
\bm{\nabla}_h \tilde{p}'' + \nu \frac{\partial^2 \tilde{\bm{u}}_l''}{\partial y^2}
\ee
where as before subscript h represents the horizontal directions $x$ and $z$. The boundary conditions are $\tilde{\bm{u}}_l''(y=0,t) = 0$ and $\partial \tilde{\bm{u}}_l''/\partial y(y\to\infty,t)=0$ with the initial condition $\tilde{\bm{u}}_l''(y,t_0) = 0$. To simplify the problem, it is useful to define a ``non-equilibrium free stream velocity'', $\tilde{\bm{u}}''_\infty$, defined as the velocity that would exist as an inviscid response to the non-equilibrium pressure gradient:
\be
\frac{\partial \tilde{\bm{u}}_\infty''}{\partial t} = 
-\frac{1}{\rho}\bm{\nabla}_h \tilde{p}'', \,\,\,\,\, \to \,\,\,\, 
\tilde{\bm{u}}_\infty''(t) = \int_{t_0}^t   -\frac{1}{\rho} \, \bm{\nabla}_h \tilde{p}'' \, dt'.
\ee
We can then use this velocity to eliminate the pressure gradient using the variable transformation $\hat{\bm{u}}(y,t)=\tilde{\bm u}_\infty''(t) - \tilde{\bm u}_l''(y,t)$. The problem then reduces to the generalized Stokes problem where the wall velocity is $\tilde{\bm u}_\infty''(t)$. From \cite{Schlichting2017} this has the solution
\begin{equation}
\hat{\bm u}(y,t) = \int_{t_0}^t \frac{\partial \tilde{\bm u}_\infty''}{\partial t} \;
\textrm{erfc} \left( \frac{y}{2\sqrt{\nu(t-t')}} \right) \; dt'.
\end{equation}
Rewriting in terms of $\tilde{\bm u}_l''$ and $\bm{\nabla}_h \tilde{p}''$ gives
\begin{equation}
\tilde{\bm u}_l''(y,t) 
= \int_{t_0}^t \left(-\frac{1}{\rho}\bm{\nabla}_h \tilde{p}''\right) \;
\textrm{erf} \left( \frac{y}{2\sqrt{\nu(t-t')}} \right) \; dt'
\end{equation}
from which the stress contribution can be obtained by differentiation, evaluation at $y=0$ and multiplication by $\nu$, and reads as follows:
\be \label{eq:neq_wallstress}
\bm{\tau}_w''(t)
= \sqrt{\nu/\pi} \int_{t_0}^t - \frac{1}{\rho}\bm{\nabla}_h \tilde{p}^{\prime \prime}(t') \,\,  (t - t')^{-1/2} dt'.
\ee
Interestingly, we can use $\tilde{\bm u}_\infty''$ to relate the non-equilibrium wall stress with the Caputo fractional derivative:
\be
\bm{\tau}_w'' = -\sqrt{\nu/\pi} \,\,\, \Gamma(1/2) \,\, D_t^{1/2} (\tilde{\bm{u}}_\infty''),
\ee
where the Caputo fractional derivative of order $\alpha$ of a signal $v(t)$ is defined as
\be
D_t^\alpha v(t) = \frac{1}{\Gamma(1-\alpha)} \, \int_{t_0}^t \frac{v^{(1)}(t')}{(t-t')^{\alpha}}\, dt'
\ee
In the equation above $0 < \alpha < 1$ is the order of the fractional derivative. The rapid wall stress model uses $\alpha = 1/2$. Relating the wall stress with the Caputo fractional derivative is useful because an efficient numerical evaluation of this type of non-local integral operator is possible, as described in \S \ref{sec:convolution_integral}.

\section{Pressure decomposition and numerical implementations}
\label{sec:implementations}

\subsection{Pressure gradient decomposition}
\label{sec:pg_decomposition}

In this section we discuss the various pressure gradient inputs to the model: $\bm{\nabla}_h P$, $\bm{\nabla}_h \overline{p}'$ and $\bm{\nabla}_h \tilde{p}''$. The first ($\bm{\nabla}_h P$) is used in evaluating the fully equilibrium fitted part to evaluate the turbulent stress as input to the LaRTE equation. $\bm{\nabla}_h \overline{p}'$ is the fluctuating pressure gradient input that directly affects the LaRTE dynamics. The last term $\bm{\nabla}_h \tilde{p}''$ is the forcing term for the non-equilibrium laminar response model described in \ref{sec:stokeslayer}.  We begin from  the pressure gradient available from LES, which corresponds to the pressure gradient horizontally filtered to the size of the LES grid, denoted by $\bm{\nabla}_h \tilde{p}$, where $\tilde{p}=p_{\rm LES}$ at $y=\Delta$. We decompose it according to these three contributions:

\be
\bm{\nabla}_h \tilde{p} = \bm{\nabla}_h P + 
\bm{\nabla}_h \overline{p}' + \bm{\nabla}_h \tilde{p}'' = 
\bm{\nabla}_h \overline{p} + \bm{\nabla}_h \tilde{p}'',
\ee
where $\bm{\nabla}_h \overline{p}=\bm{\nabla}_h P + \bm{\nabla}_h \overline{p}'$.

The ``fully-equilibrium pressure gradient'', $\bm{\nabla}_h P$ is to be used in the fitting function to model the turbulent stress at $y=\Delta$. It is obtained by temporal filtering $\bm{\nabla}_h \tilde{p}$ at a long time scale $n \,T_s$ where $n$ is some constant sufficiently greater than one and $T_s$ is the relaxation time scale of the LaRTE model. We thus write
\be
\bm{\nabla}_h P = \bm{\nabla}_h \langle \tilde{p}  \rangle_{nT_s},
\ee
where the brackets indicate one-sided exponential time filtering and the subscript denotes the corresponding filtering time-scale. The rationale for this choice is that the equilibrium time scale should be greater than the quasi-equilibrium time scale, $T_s$, such that only very slow pressure changes are included in the fitted full-equilibrium model. We chose $n=3$ as a practical compromise that works well in applications to be shown later, and results appear to be quite insensitive to this choice.

The laminar Stokes layer that develops near the wall is caused by the high-frequency components (fastest changing) pressure gradient fluctuations. Therefore we define the non-equilibrium pressure gradient input to be a high-pass temporally filtered version of the pressure gradient. This is achieved in practice by subtracting from the LES pressure gradient another low-pass filtered signal, but low-pass filtered at a high frequency. Specifically, we write
\begin{equation} \label{eq:neq_pg}
\bm{\nabla}_h \tilde{p}'' = \bm{\nabla}_h \tilde{p} - \bm{\nabla}_h \langle \tilde{p} \rangle_{t_\nu},
\end{equation}
where $\langle \cdot \rangle_{t_\nu}$ represents a temporal low-pass filter at time scale $t_\nu$. 
For $t_\nu$, the appropriate filtering time scale should be the diffusion time from the wall to the edge of the Stokes layer ($y=l_s$). We define this time scale to be $t_\nu \equiv l_s^2/\nu$. Then rewriting the Stokes layer thickness in inner units, the time scale becomes
\begin{equation} \label{eq:tnu}
t_\nu = \frac{(l_s^+)^2 \nu}{u_\tau^2}.
\end{equation}
The Stokes layer is assumed to be confined to the viscous sublayer, therefore as an approximation we let $l_s^+ \approx 12$ and $u_\tau$ is obtained from the LaRTE model.

With $\bm{\nabla}_h P$ and $\bm{\nabla}_h \tilde{p}''$ so determined, the input to the LaRTE transport equation is the ``band-pass filtered'' version of the pressure gradient equal to 

\be
\bm{\nabla}_h \overline{p}' \, = \,   \bm{\nabla}_h \langle \tilde{p} \rangle_{t_\nu} - 
\bm{\nabla}_h \langle \tilde{p} \rangle_{3T_s},
\ee
recalling that $\tilde{p} = p_{\rm LES}$ is the pressure available from LES at the first wall model point away from the wall. 

\begin{figure}
\centering
\includegraphics[width=0.8\linewidth]{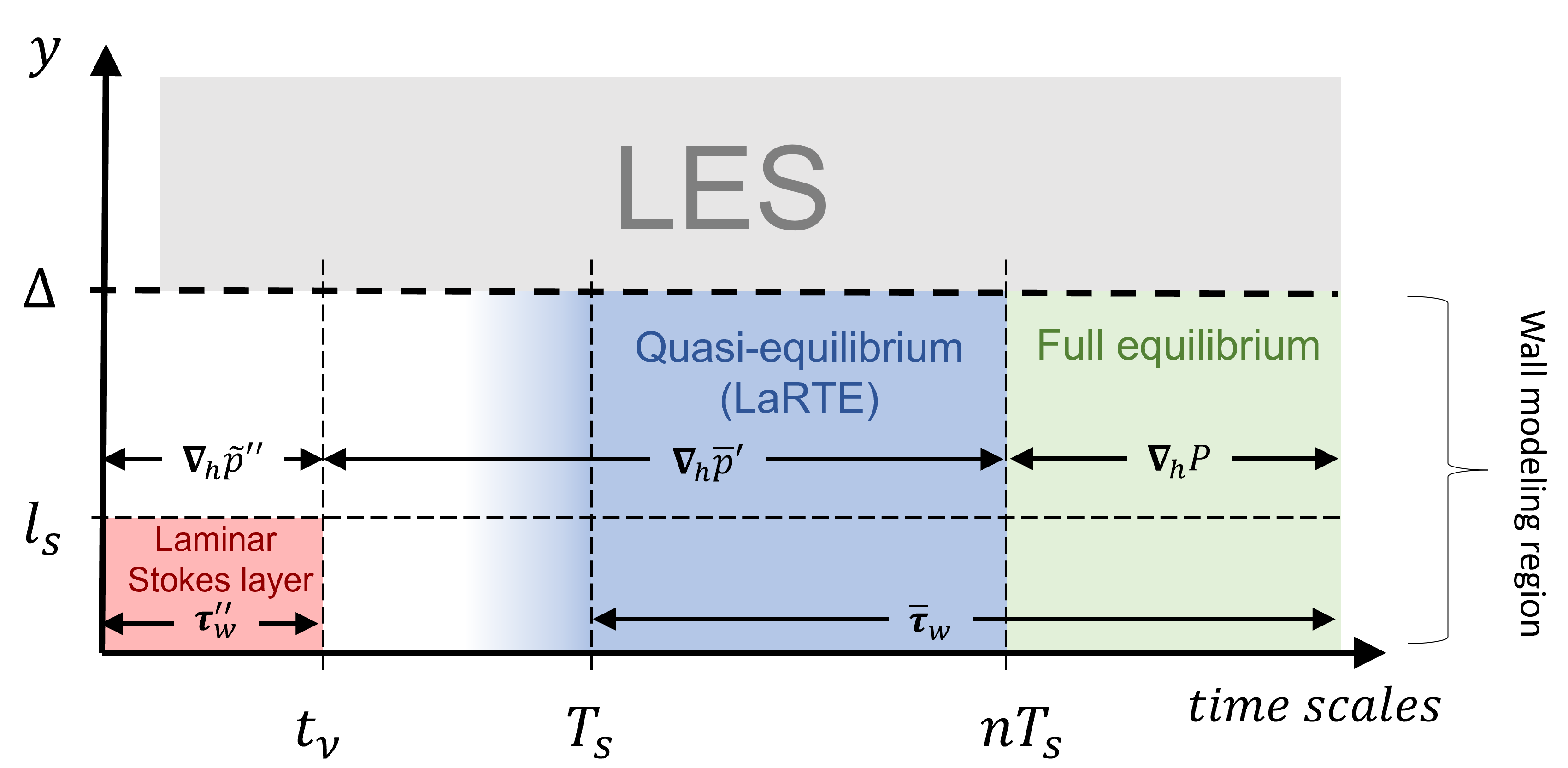}
\caption{Schematic of the various time scales and wall-normal distances considered for modeling. Different colored regions identify the corresponding wall modeling components.}
\label{fig:scale_schematic}
\end{figure}

Figure \ref{fig:scale_schematic} shows the different time scales and wall distances considered for the wall modeling region beneath $y=\Delta$. The laminar Stokes layer is confined to the fastest time scale, $t_\nu$, and smallest wall distance, $l_s$, considered. The quasi-equilibrium and full equilibrium regions on the other hand, correspond with the the largest time scales considered ($T_s$ and $nT_s$, respectively). Note that the LaRTE model has some high frequency content coming from $\bm{\nabla}_h \overline{p}'$ and $\overline{\bm{\tau}}_\Delta$, thus the blue region extends somewhat further left than $T_s$. The remaining region left in white corresponds to the turbulent portion in the wall-modeled region, below $y=\Delta$, at scales faster than $T_s$ but slower than the viscous time-scale $t_\nu$. The response of turbulence in this region (e.g. reduction of turbulent stresses due to scrambling) requires separate modeling not yet included in the present work. Note that interesting phenomena such as drag reduction due to spanwise wall oscillations or applied pressure gradients \citep{jung_mangiavacchi_akhavan_1992} would depend to a large extent on such modeling.

\subsection{Discretization of the LaRTE evolution equation}
\label{sec:Lagtimederivativenumerics}
The full model embodied by equation (\ref{eq:Lagrelax}) is a nonlinear PDE for ${\bm u}_\tau(x,y,t)$, with an elliptic diffusion term and advective term. Following the logic of the Lagrangian dynamic model implementation \citep{meneveauetal96} and acknowledging the approximate nature of various modeling assumptions to be made, we opt for efficiency over high-order numerical accuracy in the proposed numerical implementation, while aiming to maintain the main features of the model. We discretize the LaRTE evolution equation using a forward Euler method such that the friction velocity vector may be solved for explicitly. This requires evaluating all terms, including the Lagrangian time derivative, at the previous time step $t_{n-1} = t_n - \delta t$ (with $n$ the time step index) where all terms are known. We then propose to discretize the Lagrangian derivative at time $t_{n-1}$ using a semi-Lagrangian scheme \citep{staniforth_cote_1991}:
\begin{align}
\label{eq:Lagderivative}
\left[ \frac{\partial {\bm u}{_\tau}}{\partial t}  + {\bm V}_\tau \cdot \nabla_h {\bm u}{_\tau} \right](x^\prime_i,z^\prime_k,t_{n-1}) &= \left[\frac{ {d}_s {\bm u}_{\tau}}{dt \,\,}\right] (x^\prime_i,z^\prime_k,t_{n-1}) \nonumber \\
&\approx \frac{1}{\delta t} \left[ {\bm u}{_\tau}(x_i,z_k,t_n) - {\bm u}{_\tau}(x^\prime_i,z^\prime_k,t_{n-1})  \right],
\end{align}
where
$x^\prime_i = x_i-V_{\tau x} \delta t$ and 
$z^\prime_k = z_k-V_{\tau z} \delta t$ (with $i$ and $k$ position indices), and  ${\bm V}_\tau$ is evaluated from  equation \ref{eq:defVtau} using ${\bm u}_\tau(x_i,z_k,t_{n-1})$. 
 Replacing into the equation
$
  \left[ {\bm u}{_\tau}(x_i,z_k,t_n) - {\bm u}{_\tau}(x^\prime_i,z^\prime_k,t_{n-1})  \right] /\delta t = {\bf RHS}(x^\prime_i,z^\prime_k,t_{n-1}), 
$
 where ${\bf RHS}$ is the entire right-hand-side of equation \ref{eq:Lagrelax2} yields
 \be
 {\bm u}{_\tau}(x_i,z_k,t_n) = {\bm u}{_\tau}(x^\prime_i,z^\prime_k,t_{n-1}) + \delta t \, {\bf RHS}(x^\prime_i,z^\prime_k,t_{n-1}).
 \label{eq:utauRHS}
 \ee
  
The entire right-hand-side of equation \ref{eq:utauRHS} at the upstream position $(x_i-V_{\tau x} \delta t,z_k-V_{\tau z} \delta t)$ at the   time $t_{n-1}$ is obtained using first-order bilinear spatial interpolation of the grid values on the plane, as was done in 3D in \cite{meneveauetal96}. The additional numerical diffusion associated with the low-order interpolation reduces the need to include the horizontal diffusion term ${\bm \nabla}_h \cdot {\bm D}_\tau$ which would require additional modeling and numerical cost associated with solving an elliptic problem. Thus, we neglect the term ${\bm \nabla}_h \cdot {\bm D}_\tau$ altogether in practical implementations in our LES (however, see Appendix \ref{app:diffusion} for explicit form of one part of this term).
Finally, evaluation of ${\rm \bm RHS}$  requires the $\partial_t \bm s$ term. It is  discretized using backward differencing, as
$\partial_t \bm s|_{n-1} = (\bm s|_{n-1}-
\bm s|_{n-2})/\delta t$ all evaluated at the interpolated position $(x_i-V_{\tau x} \delta t,z_k-V_{\tau z} \delta t)$.

\subsection{Evaluating the temporal convolution integral}
\label{sec:convolution_integral}

\cite{jiang2017soe}  developed a method for ``fast evaluation of the Caputo fractional derivative" which significantly reduces storage and computational cost requirements thus making the computation of the convolution integral practical. To summarize, their method decomposes the integral into a local and history parts where the history contribution is evaluated efficiently by making a sum-of-exponentials approximation to the kernel. An exponential kernel has the advantage that the current value of the convolution depends only on the previous time-step value of the convolution and a local term, as exploited in many applications where time filtering is needed (as e.g. in \cite{meneveauetal96} and in other instances of exponential time filtering applied in this paper).  

Since the sum-of-exponentials approximation algorithm is critical for the model, we will describe here the basic details of it pertaining to our application with $\alpha = 1/2$. Our task is to find an efficient way of computing the convolution integral in equation \ref{eq:neq_wallstress}. To simplify notation we let $\bm{G}(t) \equiv -\rho^{-1}\bm{\nabla}_h \tilde{p}''(t)$. Then the non-equilibrium wall stress is given by
\begin{equation}
\label{eq:neq_model_discrete}
\bm{\tau}_w''(t_n) = \sqrt{\nu/\pi} \int_{t_0}^{t_n} \bm{G}(t')(t_n-t')^{-1/2} \; dt'
\end{equation} 
where $t_n$ is the current time and $n$ is, as before, the time step index. The sum-of-exponentials (SOE) approximation for the kernel reads:
\begin{equation} \label{eq:soe_approx}
(t_n - t')^{-1/2} \approx \sum_{m=1}^{N_{exp}} \omega_m e^{-s_m (t_n - t')}
\end{equation}
where the constants $\omega_m$ and $s_m$ are determined a priori as a function of the time-step size for the SOE approximation, $\delta t$, the time duration considered, $T$, and the desired maximum error for the SOE approximation of the kernel, $\epsilon$. According to \cite{jiang2017soe}, the number of exponential terms, $N_{\exp}$, is also a function of these parameters and can be estimated by the expression
\begin{equation} \label{eq:Nexp}
N_{\exp }=\mathcal{O}\left(\log \frac{1}{\epsilon}\left(\log \log \frac{1}{\epsilon}+\log \frac{T}{\delta t}\right)+\log \frac{1}{\delta t}\left(\log \log \frac{1}{\epsilon}+\log \frac{1}{\delta t}\right)\right).
\end{equation}
 $\epsilon$ is the error associated with the approximation in equation \ref{eq:soe_approx} (not to be confused with the error from discretizing equation \ref{eq:neq_model_discrete}). 
 The simulations in this paper will mostly use $\delta t \sim 4\times 10^{-4}$ and so the constants were computed using $\delta t=4\times 10^{-4}$. Also, it was found that in order to guarantee good accuracy in all cases considered, we required $\epsilon=10^{-9}$. We used $T=1$ although this parameter was seen to affect the coefficients very little as long as $T \gg \delta t$. The optimization approach by \cite{jiang2017soe} yields $N_{\exp}=48$ although fewer terms (obtained by using larger $\epsilon$) could be used while still yielding reasonable accuracy. Appendix \ref{app:soe} provides information about the computed constants as well as a more detailed verification of the numerical method.

The integral is divided into local and history parts (in time)
\begin{equation} \label{eq:neq_decomposition}
\bm{\tau}_w'' = \bm{\tau}_{w,l}'' + \bm{\tau}_{w,h}''
\end{equation}
where
\begin{align}
\bm{\tau}_{w,l}''(t_n) &\equiv \sqrt{\nu/\pi} \int_{t_{n-1}}^{t_n} \bm{G}(t')(t_n-t')^{-1/2} \; dt' \nonumber \\
\bm{\tau}_{w,h}''(t_n) &\equiv \sqrt{\nu/\pi} \int_{t_0}^{t_{n-1}} \bm{G}(t')(t_n-t')^{-1/2} \; dt'. 
\end{align}
The local part is evaluated using the ``L1 method'' \citep{li_zeng_2015}:
\begin{equation} \label{eq:neq_local}
\bm{\tau}_{w,l}'' \approx 2 \bm{G}(t_{n-1/2}) \sqrt{\frac{\nu \Delta t_n}{\pi}}
\end{equation}
where $\Delta t_n = t_n - t_{n-1}$ and $\bm{G}(t_{n-1/2}) = 0.5(\bm{G}(t_{n})+\bm{G}(t_{n-1}))$. The history part is evaluated by replacing the kernel with a sum-of-exponential approximation from equation \ref{eq:soe_approx}. This sum-of-exponentials approximation is useful because it allows the integral to be computed recursively. The history term can then be computed using
\begin{equation} \label{eq:neq_history}
\bm{\tau}_{w,h}''
\approx \sqrt{\nu/\pi} \sum_{m=1}^{N_{exp}} \omega_m \bm{I}_m (t_n)
\end{equation}
where
\begin{align} \label{eq:neq_I}
\bm{I}_m(t_n) &= \int_{t_0}^{t_{n-1}} \bm{G}(t')e^{-s_m(t_n-t')}dt' \nonumber \\
&= e^{-s_m(t_n-t_{n-1})} \bm{I}_m(t_{n-1}) + \int_{t_{n-2}}^{t_{n-1}} \bm{G}(t')e^{-s_m(t_n-t')}dt' \nonumber \\
&\approx e^{-s_m \Delta t_n} \left[ \bm{I}_m(t_{n-1}) + \frac{\bm{G}(t_{n-3/2})}{s_m} (1-e^{-s_m \Delta t_{n-1}}) \right].
\end{align}
The total non-equilibrium wall stress can then be computed using equation \ref{eq:neq_decomposition} together with equations \ref{eq:neq_local}, \ref{eq:neq_history}, and \ref{eq:neq_I}. The advantage of this method is that it requires $O(N_{exp})$ storage and $O(N_T N_{exp})$ computational work whereas a direct method requires storing the entire time evolution, i.e. $O(N_T)$ storage and $O(N_T^2)$ work which becomes unwieldy for long simulations.

\section{Tests in equilibrium and non-equilibrium channel flow}
\label{sec:channel}

To test the new wall model (with both quasi-equilibrium LaRTE and non-equilibrium components), large eddy simulations are conducted  for statistically stationary channel flow as well as for channel flow with a large step change in the spanwise pressure gradient (referred to as sudden spanwise pressure gradient or SSPG).  

\subsection{Statistically stationary channel flow}
\label{sec:steady_state}
First, the LaRTE wall model together with the non-equilibrium part is implemented in a simulation of statistically steady state channel flow at various Reynolds numbers. This is a flow in which the traditional equilibrium wall model  typically provides good results. The objective is thus mainly to ensure that similarly good results are obtained using the new model as well as to document its various features, such as typical orders of magnitudes of the terms appearing in the Lagrangian relaxation transport equation for the friction-velocity vector. Simulations use LESGO, an open-source, parallel, mixed pseudo-spectral and centered finite difference LES code available on Github \citep{LESGOsite}. The Lagrangian scale-dependent dynamic subgrid stress model \citep{bou05} is used in the bulk of the flow. The near-wall region is modeled using the new wall models proposed here: the LaRTE model governed by equation \ref{eq:Lagrelax} (with closure and simplifications according to equation \ref{eq:Lagrelax2}) and the laminar non-equilibrium model governed by equation \ref{eq:neq_wallstress}. A wall stress boundary condition is applied consisting of the superposition between the two models (i.e. $\tilde{\bm\tau}_w = \overline{\bm\tau}_w + \bm{\tau}_w''$).
Further notes regarding implementation are discussed in \S \ref{sec:implementations} and fits needed for the LaRTE model are provided in Appendix C.

First, simulations are performed with  friction Reynolds numbers based on the half channel height of $Re_\tau = $ 1{,}000 and 5{,}200. The domain size, number of grid points, and grid size are $(L_x, L_y, L_z)/h = (8\pi,2,3\pi)$, $(N_x, N_y, N_z) = (128,30,48)$, and $(\Delta_x, \Delta_y, \Delta_z)/h = (0.196,0.067,0.196)$, respectively. In inner units the grid size for $Re_\tau = $1{,}000 and 5{,}200 are $(\Delta_x^+, \Delta_y^+, \Delta_z^+) = (196,67,196)$ and $(\Delta_x^+, \Delta_y^+, \Delta_z^+) = (1021,347,1021)$, respectively. Several additional simulations are performed at even higher Reynolds numbers ($Re_\tau = \{0.2,\, 1,\ 5\}\times 10^5$) using the same number of grid points in order to ensure applicability at arbitrarily high Reynolds numbers.
As can be seen these are very coarse WMLES, very different from the much finer resolutions required for WRLES. 

In LESGO the wall model takes information from the first grid point away from the wall (i.e. $\Delta = \Delta_y/2$). The wall model heights for all friction Reynolds numbers considered are summarized in table \ref{table:model_height}. These wall-model heights lie within the log-layer. The proposed new wall model
is applied using the LES data at $y=\Delta$. A ``$2\Delta$ spatial filter'', like that used in \cite{bou05}, is applied to the LES velocity at $y=\Delta$ which is provided as the velocity input to equation \ref{retd_fit} to model the turbulent stress at that position. This is primarily done to reduce log-layer mismatch \citep{Yang2017logmismatch} without causing an excessively sluggish response in the wall stress which would occur if the velocity was time filtered instead. The pressure gradient, on the other hand, is not spatially filtered but instead is temporally filtered with the single-sided exponential filter with the decomposition and filtering time scales described in \S \ref{sec:pg_decomposition}.

\begin{table}
\centering
\begin{tabular}{c|c}
$Re_\tau$  & $\Delta^+$ \\ 
\noalign{\hrule}
$1.0\times 10^3$ & 33 \\
$5.2\times 10^3$ & 173 \\
$2.0\times 10^4$ & 667 \\
$1.0\times 10^5$ & 3333 \\
$5.0\times 10^5$ & 16667 \\
$3.17\times 10^3$ & 106 \\
\end{tabular}
\caption{Wall model height in inner units $\Delta^+$ for all friction Reynolds numbers simulated. $Re_\tau = 3170$ corresponds with the steady state friction Reynolds number long after the application of the spanwise pressure gradient presented in \S \ref{sec:spanwise}.}
\label{table:model_height}
\end{table}

\begin{figure}
\centering
\includegraphics[width=.5\linewidth]{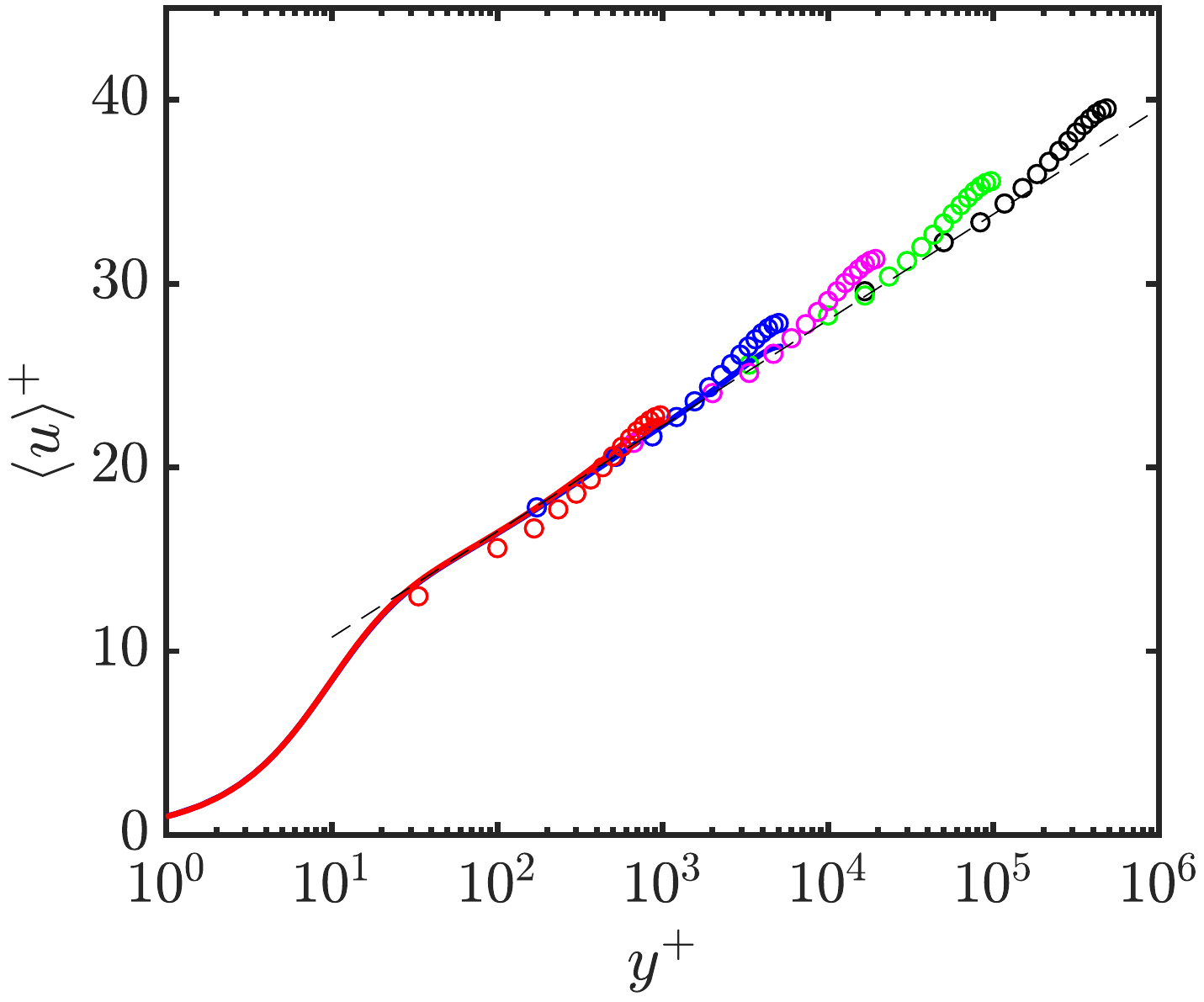}
\caption{Open circles: mean velocity profiles from WMLES using the LaRTE and non-equilibrium wall model for $Re_\tau =$ 1{,}000 (red), 5{,}200 (blue), 20{,}000 (magenta), $10^5$ (green) and $5\times 10^5$ (black). Lines: DNS from Lee and Moser (2015) at $Re_\tau =$ 1{,}000 (red line), 5{,}200 (blue line), and log-law $\langle u \rangle^+=\ln(y^+)/0.4+5.0$ (dashed line).}
\label{fig:mean_velocity_profile}
\end{figure}

Figure \ref{fig:mean_velocity_profile} displays the mean velocity for the five  Reynolds numbers tested. DNS results for the two lower Reynolds number cases are also included. All profiles follow the expected law-of-the-wall and the results confirm that the new wall model, with combined LaRTE and non-equilibrium parts, is able to reproduce results quite comparable to those of the standard equilibrium wall model. Note that there is a slight log-layer undershoot for $Re_\tau = $ 1{,}000 and an overshoot of the profile 
in the wake-region at the center of the channel. Similar trends have been obtained in WMLES using different codes and SGS models \citep{yang2015integral}, and are likely attributable to the SGS modeling details and not the wall model. As shown in the recent wall-model independent analysis by \cite{Lozano-Duran_Bae_2019}, LES accuracy in the outer region of wall bounded flows is highly sensitive to details of the ratio of grid resolution compared to outer length-scale (rather than Reynolds number). Examinations of second order moments of velocity fluctuations (not shown) yield results very similar to those in \cite{bou05}. We conclude that the new wall model generates baseline velocity statistics that are similar to those obtained from the standard equilibrium wall model. 

\begin{figure}
\centering
\includegraphics[width=\linewidth]{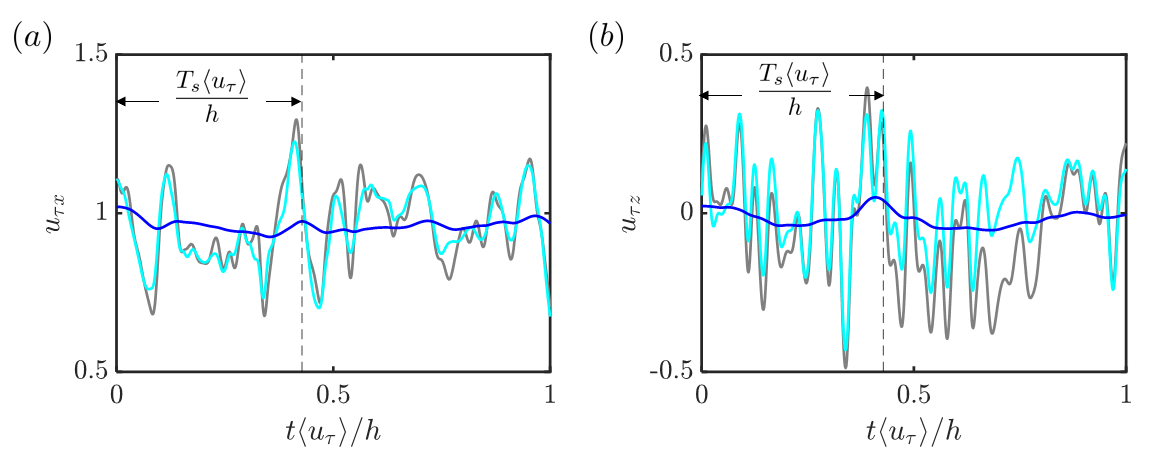}
\caption{Time signals of relevant terms in the LaRTE model at some arbitrary representative point at the wall from LES or channel flow at $Re_\tau$=1{,}000 and $\Delta/h=1/30$. Time signals shown are for the terms $\overline{\bm \tau}_\Delta/u_\tau$  (\protect\thickcyanline), $(-\Delta {\bm \nabla}_h \overline{p}/\rho + \overline{\bm \tau}_\Delta)/u_\tau$  (\protect\thickgrayline) and 
$ {\bm u}_\tau$  (\protect\thickblueline). Panel (a) shows the $x$-component and (b) the $z$-component terms. The vertical dashed line shows the relaxation time scale $T_s \langle u_\tau \rangle/h \approx 0.45$.}
\label{fig:relaxation_terms}
\end{figure}

Next, we illustrate by means of time signals at a representative point on the wall the various terms in equation \ref{eq:Lagrelax} that is being solved at each point following the implementation described in \S \ref{sec:Lagtimederivativenumerics}. In figure \ref{fig:relaxation_terms} we show, in cyan, signals of the input stress vector $\overline{\bm{\tau}}_\Delta$ at $y=\Delta$, evaluated using the fitted equilibrium model in equation \ref{retd_fit}. The stress is divided by $u_\tau$, the magnitude of the obtained friction velocity. The gray line shows the same, but with the horizontal pressure gradient added, the quantity towards which the friction velocity vector ${\bm u}_\tau$ relaxes, with relaxation time scale $T_s$. The cyan and gray lines are generally close, showing that the effect of the pressure gradient is smaller but not negligible compared to the imposed turbulent stress.   The blue line in figure \ref{fig:relaxation_terms} shows the friction velocity resulting from the LaRTE solution. In this flow, the characteristic mean value of $T_s$ can be estimated as $T_s \langle u_\tau \rangle / h = (\Delta/h) f(\Delta^+) = (1/30) f(1000/30) \approx 0.45$. As is evident, major fluctuations of ${\bm u}_\tau$ occurring at time scales smaller than $T_s$ have been filtered out almost entirely. Only low frequency variability is left, internally consistent with the notion of quasi-equilibrium that underlies the assumption of the profile scaling in inner units. Note that if an equilibrium model were used the wall stress would fluctuate at levels comparable with the cyan signal.
 
\begin{figure}
\centering
\includegraphics[width=\linewidth]{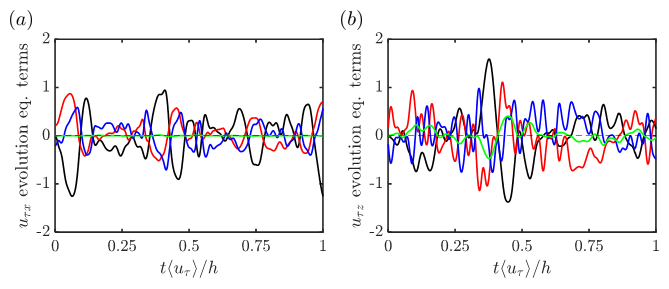}
\caption{Time signals of terms in the evolution equation for ${\bm u}_\tau$, Eq. \ref{eq:Lagrelax} at some arbitrary representative point at the wall from LES or channel flow at $Re_\tau$=1{,}000 and $\Delta/h=1/30$. Time signals shown are for the terms $\partial \bm{u}_\tau/\partial t$  (\protect\thickblackline),
${\bm V}_\tau \cdot {\bm \nabla}_h {\bm u}_\tau$  (\protect\thickredline),
$-T_s^{-1}\left[u_\tau^{-1}(-\Delta {\bm \nabla}_h \overline{p}/\rho + \overline{\bm \tau}_\Delta)-{\bm u}_\tau \right]$  (\protect\thickblueline), and
$-u_\tau(\delta^*_\Delta/\Delta) \partial \overline{\bm s}/\partial t$  (\protect\thickgreenline).}
\label{fig:Lag_rewm_terms_compare}
\end{figure}

Next, signals of the individual terms in Eq. \ref{eq:Lagrelax} are presented in figure \ref{fig:Lag_rewm_terms_compare}. The Eulerian time derivative shown in black displays some anticorrelated trend with the advective term shown in red. This is expected for transported quantities, and leads to smaller magnitudes of the Lagrangian time derivative as compared to the Eulerian time derivative. The blue line shows the entire relaxation towards equilibrium term which essentially drives the rate of change of the friction velocity vector. The non-standard term with the Eulerian time derivative of the orientation vector ${\bm s}$ is negligible in the $x$-direction while it shows some contribution in the spanwise direction.

\begin{figure}
\centering
\includegraphics[width=\linewidth]{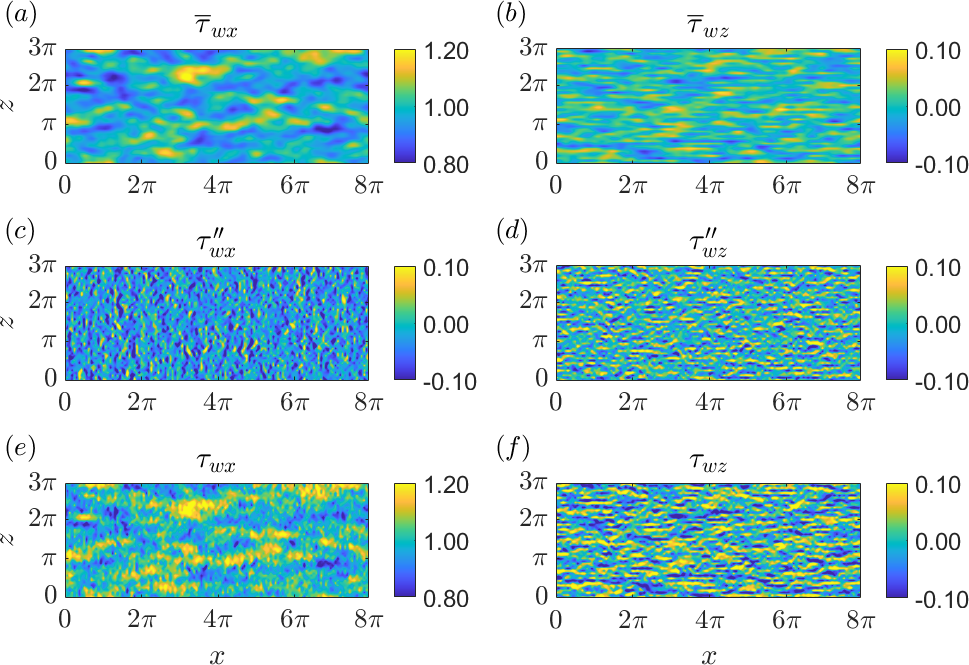}
\caption{Snapshots of the wall stress for $Re_\tau=$1{,}000. The top row shows the quasi-equilibrium stress from the LaRTE model for both streamwise and spanwise components, the middle row shows the laminar layer non-equilibrium portion, and the bottom the total (sum of both).}
\label{fig:contours_ret1000}
\end{figure}

Wall stress contours for $Re_\tau=$ 1{,}000 are presented in figure \ref{fig:contours_ret1000} for a single snapshot of one of the LES realizations. As can be seen from the top row, the LaRTE quasi-equilibrium stress shows elongated structures that extend over relatively long distances downstream. The fluctuations occur, as expected around a value of $\overline{\tau}_{wx} \approx 1$. The spanwise stress component $\overline{\tau}_{wz}$ has zero mean and fluctuations that appear to occur at smaller scales, generally consistent with elongated structures that display larger variability in the transverse direction than in the streamwise direction.  The middle row shows the contribution from the laminar non-equilibrium portion of the model. In spite of the backward time integration that should smooth signals to some degree, these fields display much smaller-scale fluctuations. These reflect fluctuations in pressure gradients in both streamwise and spanwise directions that tend to occur at scales similar to the LES grid scale. The bottom row shows contours of the sum of both contributions, combining the streamwise elongated structure and the smaller-scale fluctuations from the laminar non-equilibrium part of the model. 

\begin{figure}
\centering
\includegraphics[width=\linewidth]{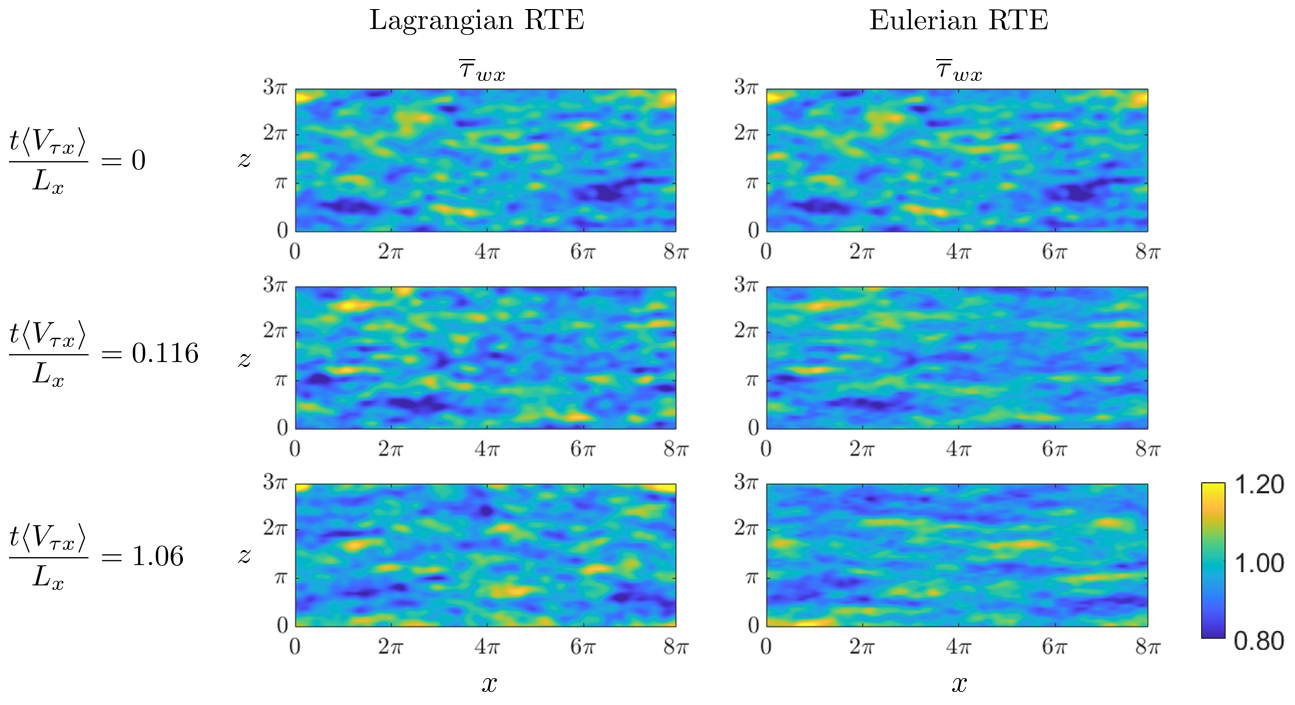}
\caption{Streamwise wall stress, $\overline{\tau}_{wx}$, contours for several time instances for $Re_\tau = 1,000$. Compares Lagrangian RTE (left column) with Eulerian RTE (right column) where the advection term is excluded. Both models are initialized with the same data as shown in the top row. Time is non-dimensionalized with $\langle V_{\tau x} \rangle \approx 7.93 \, \langle u_\tau \rangle$ and $L_x=8\pi \, h$.}
\label{fig:lagrangian_vs_eulerian}
\end{figure}

It is of interest to explore further the qualitative differences between an Eulerian and a Lagrangian time derivative in applying the LaRTE model. To this effect, we select some time during the LES using the LaRTE approach and denote that time as $t=0$. Then we continue the LES using the Lagrangian version of LaRTE and perform another simulation that continues using the Eulerian version, i.e. simply omitting the advective derivative ${\bm V}_\tau \cdot {\bm \nabla}_h {\bm u}_\tau$ from the evolution equation. Figure \ref{fig:lagrangian_vs_eulerian} shows the results in the form of contour plots of the x-component of the modeled wall stress, $\overline{\tau}_{wx}=u_\tau u_{\tau x}$. By construction they both agree at $t=0$ but begin to differ at later times, significantly. As confirmed by examining animations, the Eulerian version ``pins'' fluctuations at the wall while perturbations from imposed stress at $y=\Delta$ travel downstream. The time filtering implicit in the relaxation equation then ``smears'' and elongates the structures excessively in the streamwise direction. In the Lagrangian version shown to the left, perturbations are allowed to travel downstream, including the time-filtered versions that therefore maintain their more compact integrity as time progresses. We conclude that the Lagrangian version appears more physically reasonable. We remark, however, that we have no ``true'' distribution (e.g. from DNS) to compare with, since we would need to evaluate either Eulerian and Lagrangian time averaging from the DNS, and similar differences would be obtained, without necessarily indicating which one is ``better'' or ``true''. Having seen significant differences in predicted stress distribution between Eulerian and Lagrangian versions of the model, and the latter being directly motivated by the underlying integral momentum equation, we continue using the Lagrangian version for the rest of this paper.

\begin{figure}
\centering
\hspace{2cm}
\includegraphics[width=\linewidth]{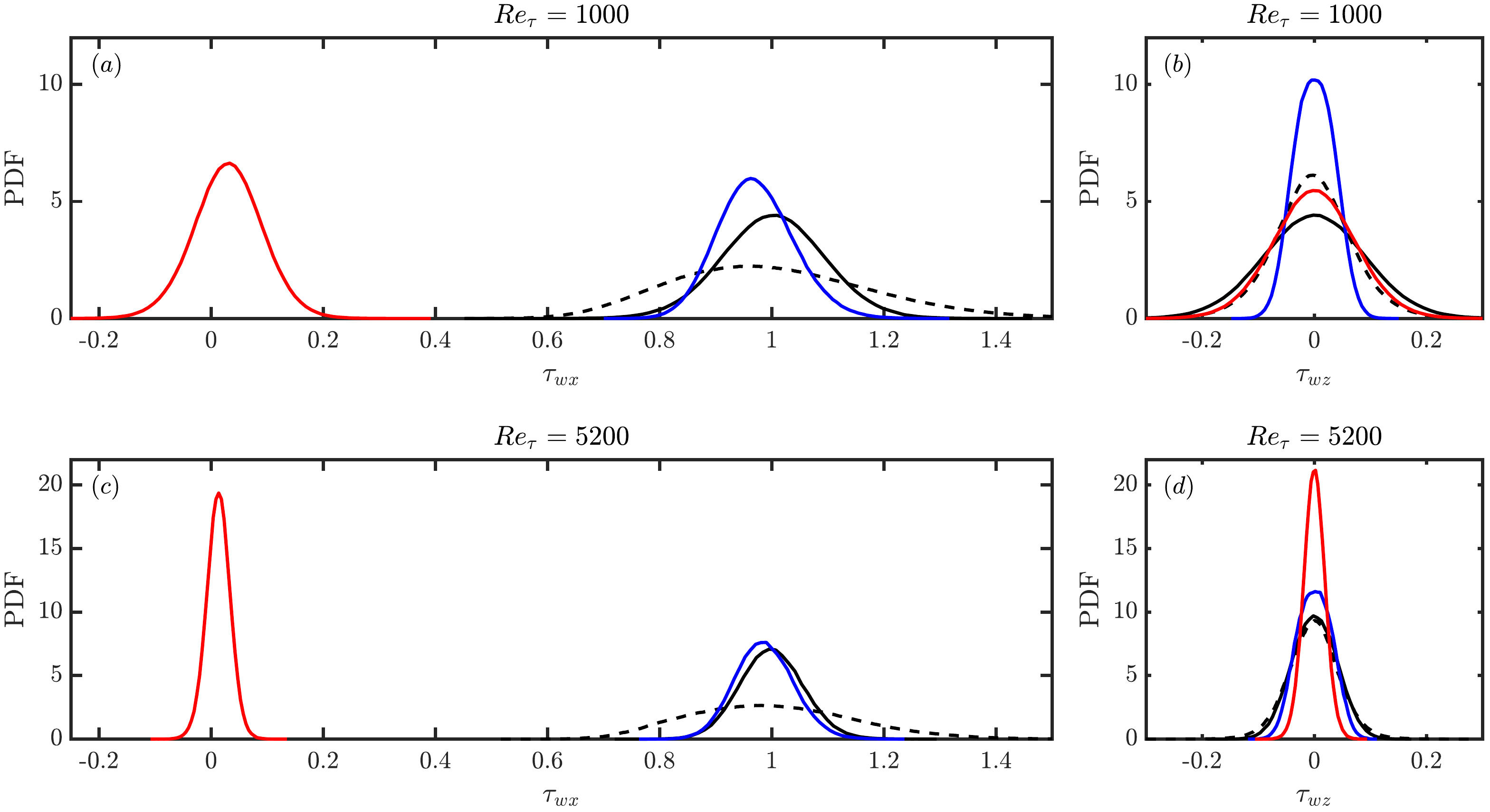}
\caption{PDFs for $\tau_{wx}$ and $\tau_{wz}$; $(a,b)$ $Re_\tau = 1000$; $(c,d)$ $Re_\tau = 5200$. The PDF curves correspond with the filtered DNS (\protect\thickblackdashed), the LaRTE model (\protect\thickblueline), the non-equilibrium model (\protect\thickredline), and the composite model (LaRTE + non-equilibrium) (\protect\thickblackline). DNS data obtained from the Johns Hopkins Turbulence Database \cite{JHTDBwebsite,graham2016web}. The DNS PDFs are obtained from the Gaussian filtered wall stress where the filtering size is the same as the LES mesh size in the horizontal directions. }
\label{fig:pdfs}
\end{figure}

More quantitative characterization of the stress fluctuations is provided by the probability density function (PDF) of each component of the wall stress. The PDFs obtained from the 
$Re_\tau=$ 1{,}000 and 5{,}200 could be compared to filtered DNS data at the same Reynolds numbers. LES data were collected over five separate uncorrelated simulations to obtain better convergence of statistics. DNS data were obtained from a public database   \citep{JHTDBwebsite} and the instantaneous local wall stress was spatially filtered horizontally using a Gaussian filter at the same scale as the LES grid. The PDFs are shown in figure \ref{fig:pdfs}. The PDF from the filtered DNS (dashed line) peaks around  $\overline{\tau}_{wx}=1$ and $\overline{\tau}_{wz}=0$. The quasi-equilibrium (LaRTE) part of the model (blue lines) peaks at the same expected values, but display significantly narrower distributions owing to the time filtering that reduces the fluctuations consistent with the notion of quasi-equilibrium. The laminar Stokes layer model that only models the fast laminar response in the viscous sublayer provides additional fluctuations. However, for the streamwise directions, these fluctuations are of smaller magnitude than those for the filtered DNS. This shows that the model is still missing significant parts of the streamwise stress fluctuations. Additional modeling is likely needed to account for these additional fluctuations that belong neither to the quasi-equilibrium nor the rapid laminar sublayer response parts of the dynamics. For example, a recent statistics-constrained machine learning approach \citep{Haoetal2021} may provide further improvements for predicting this missing part of the wall stress.  We note that in the spanwise direction, the PDFs agree better, in fact slightly overestimating the fluctuations for the $Re_\tau = $ 1{,}000 case but predicting the spanwise fluctuations PDF for the $Re_\tau = $ 5{,}200 case very well. From figure \ref{fig:pdfs} we can also see that as the Reynolds number increases, the PDFs of the non-equilibrium components (blue curves) narrow. As can be expected from equation \ref{eq:neq_wallstress} that shows the laminar non-equilibrium portion of the stress to be proportional to $\nu^{1/2}$, the stress contribution from the laminar Stokes layer near the wall in fact vanishes in the limit of infinite Reynolds number, unlike fluctuations expected to occur due to turbulence in the wall layer. These contributions are not included in the current model and must await further developments outside of the scope of the present paper. 
 
 We also note that when using the single-sided exponential filter with a fluctuating and short filtering time scale such as $t_\nu$, some undesirable trends can be obtained such as that the mean (in space or time) of a variable may not be exactly equal to the mean of the filtered variable. Because of this last detail, the PDF of the non-equilibrium model has a non-zero mean as seen in figure \ref{fig:pdfs}. This should be kept in mind whenever using the temporal exponential filter with a time-dependent averaging time-scale.

\subsection{Channel flow with sudden spanwise pressure gradient (SSPG)}
\label{sec:spanwise}

Next we discuss a highly non-equilibrium test case
following the work of \cite{lozano2020non}. A large spanwise pressure gradient, $\partial p_\infty/\partial z$, is applied to a statistically steady turbulent channel base flow after $t=0$. Particularly, we follow the case presented in their wall modeling results section in which the initial flow ($t=0$) is standard channel flow with $Re_\tau=$1{,}000 after which ($t>0$) a spanwise pressure gradient is suddenly applied with strength $\partial p_\infty/\partial z=10 \,\partial p_\infty/\partial x = 10 \rho u_{\tau 0}^2/h$ (where $u_{\tau 0}$ is the mean friction velocity of the initial condition and $h$ the channel half-height). The flow is initialized with the results from \S \ref{sec:steady_state}. The results presented in this section use the same code with the same mesh, subgrid scale and wall model, etc.. We should note that dynamic time stepping is used in order to maintain a constant CFL. The time step size stays within the range $1\times 10^{-4} \leq \delta t u_{\tau0}/h \leq 4 \times 10^{-4}$ from steady state to long after the application of the SSPG.

\begin{figure}
\centering
\includegraphics[width=\linewidth]{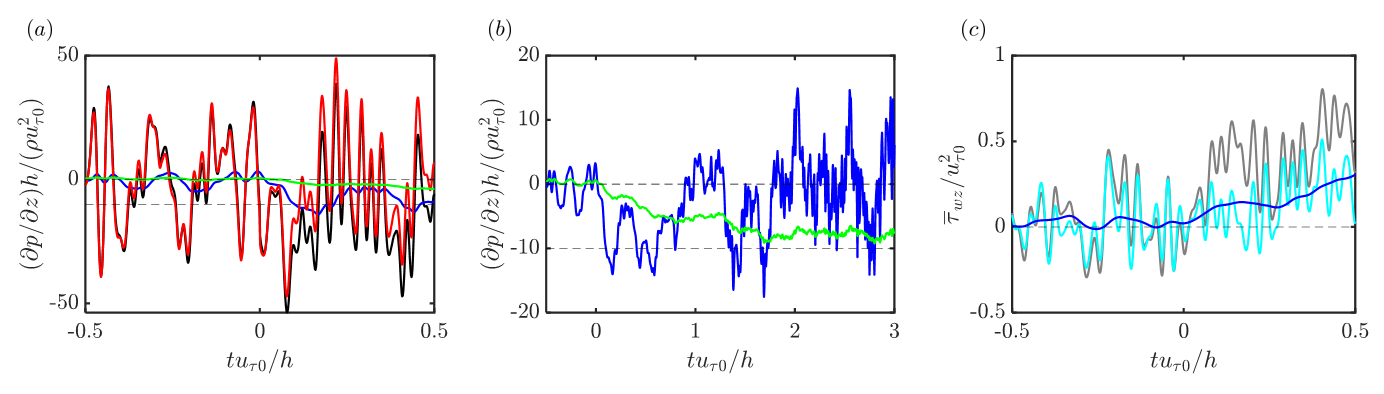}
\caption{Time signals at some arbitrary representative horizontal point for (a,b) spanwise pressure gradient components and (c) spanwise quantities in the LaRTE model relevant for $\overline{\tau}_{wz}$. (a,b): LES pressure gradient $\partial \tilde{p}/\partial z$ (\protect\thickblackline), non-equilibrium pressure gradient $\partial p'' /\partial z$ (\protect\thickredline), band-pass filtered pressure gradient $\partial \overline{p}'/\partial z$ (\protect\thickblueline), and equilibrium pressure gradient $\partial P/\partial z$ (\protect\thickgreenline) all normalized with $h/(\rho u_{\tau 0}^2)$. (c): quasi-equilibrium spanwise wall stress $\overline{\tau}_{wz}$ (\protect\thickblueline), $-\Delta (\partial \overline{p}/\partial z)/\rho + \overline{\tau}_{\Delta z}$ (\protect\thickgrayline), and $\overline{\tau}_{\Delta z}$ (\protect\thickcyanline) all normalized with $u_{\tau 0}^2$. Thin dashed horizontal lines indicate steady-state values before and after the SSPG.}
\label{fig:time_signals_sspg}
\end{figure}

First, in figure \ref{fig:time_signals_sspg}(a,b) we show pressure gradient signals at an arbitrary point corresponding to the LES pressure gradient input $\partial \tilde{p}/\partial z$ (black line), and its three constituent parts consistent with the discussion of \S \ref{sec:pg_decomposition}: the long-time average pressure gradient $\partial P/\partial z$ (green line) entering into the full equilibrium fitted model, the band-pass filtered  fluctuating pressure gradient
$\partial \overline{p}'/\partial z$ (blue line) that enters the quasi-equilibrium LaRTE equation, and the rapid non-equilibrium
$\partial \tilde{p}''/\partial z$ (red line) that affects mostly the viscous sublayer if sufficiently fast. As is evident in figure \ref{fig:time_signals_sspg}(a), $\partial \tilde{p}''/\partial z$ captures the majority of the LES pressure gradient fluctuations, $\partial P/\partial z$ captures only the ``equilibrium'' or very slowly varying pressure gradient, and $\partial \overline{p}'/\partial z$ captures any remaining fluctuations.  Figure \ref{fig:time_signals_sspg}(b) shows more clearly that at the onset of the SSPG (t=0) the equilibrium pressure gradient slowly relaxes to its new steady state value and that the strength of the quasi-equilibrium pressure gradient fluctuations grows. Both of these pressure gradient signals are inputs to the LaRTE model whose wall stress and relevant relaxation terms are shown in figure \ref{fig:time_signals_sspg}(c). Here we can see the importance of the quasi-equilibrium pressure gradient in the LaRTE model grows upon application of the SSPG.

\begin{figure}
\centering
\includegraphics[width=0.9\linewidth]{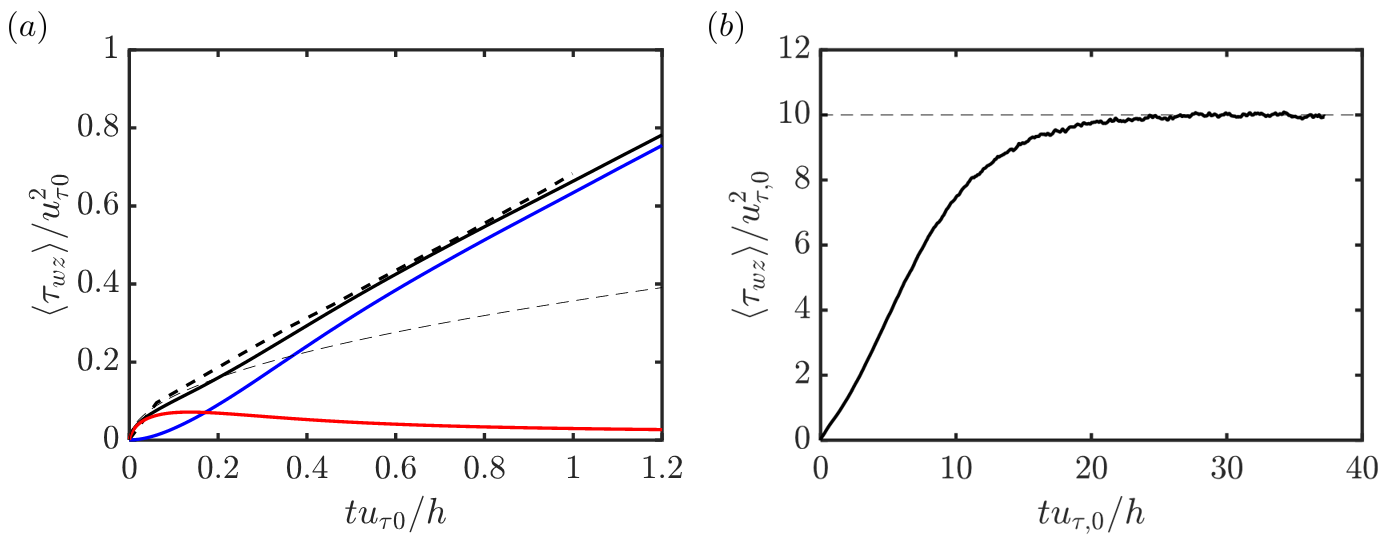}
\caption{Spanwise wall stress after sudden spanwise pressure gradient (a) after a short period and (b) after a long period. (a): DNS from \cite{lozano2020non} (\protect\thickblackdashed), composite wall stress $\langle \overline{\tau}_{wz} \rangle + \langle \tau_{wz}'' \rangle$ (\protect\thickblackline), quasi-equilibrium wall stress $\langle \overline{\tau}_{wz} \rangle$ (\protect\thickblueline), non-equilibrium wall stress $\langle \tau_{wz}'' \rangle$ (\protect\thickredline), and laminar solution for Stokes's first problem (\protect\thinblackdashed). Angled brackets indicates ensemble averaging over the horizontal plane and five separate simulations for (a) and ensemble averaging over the horizontal plane for (b). $Re_{\tau 0}=1000$ and $\partial p_\infty/\partial z = 10 \partial p_\infty/\partial x$ for $t>0$.}
\label{fig:sspg_twz_ret1000}
\end{figure}

Next we present the plane-averaged wall stress response to the SSPG.
Figure \ref{fig:sspg_twz_ret1000} shows the spanwise wall stress after the SSPG has been applied compared with the DNS of \cite{lozano2020non}. Panel (a) shows the wall stress decomposition after the initial transient and (b) shows the wall stress behavior long after the SSPG was applied. The trends are in agreement with expectations. For a brief time ($0 \leq t u_{\tau 0}/h \leq 0.05$) the wall stress follows the laminar solution closely, during which the non-equilibrium component is dominant compared with the quasi-equilibrium component. Afterwards the balance is reversed. Note that the LaRTE model responds faster than the relaxation time scale which is $T_s u_{\tau 0}/h \approx 0.45$. This is due to the inclusion of the band-pass filtered pressure gradient, $\bm{\nabla}_h \overline{p}'$, in the LaRTE model. Without this pressure gradient, $\overline{\tau}_{wz}$ is delayed by a time of order $T_s$. On the contrary, if no high-pass filtering is done, $\bm{\nabla}_h p''= 0$, $\overline{\tau}_{wz}$ is nearly linear initially and unable to capture the $\sqrt{t}$ trend corresponding to the laminar Stokes layer. Therefore, low-pass filtering is needed to prevent the overly sluggish behavior of the quasi-equilibrium model and high-pass filtering plus the inclusion of the laminar non-equilibrium model is needed to get the correct $\sqrt{t}$ behavior initially.

\begin{figure}
\centering
\includegraphics[width=0.9\linewidth]{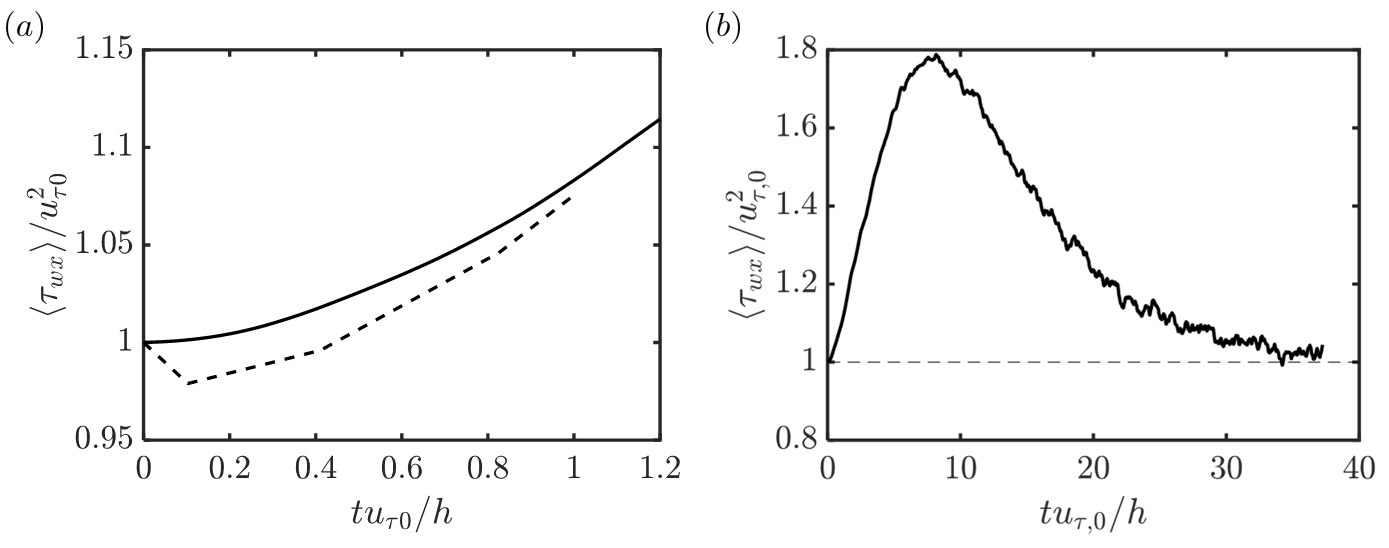}
\caption{Streamwise wall stress after sudden spanwise pressure gradient (a) after a short period and (b) after a long period. (a): DNS from \cite{lozano2020non} (\protect\thickblackdashed) and composite wall stress $\langle \overline{\tau}_{wx} \rangle + \langle \tau_{wx}'' \rangle$ (\protect\thickblackline). Angled brackets indicates ensemble averaging over the horizontal plane and five separate simulations for (a) and ensemble averaging over the horizontal plane for (b). $Re_{\tau 0}=1000$ and $\partial p_\infty/\partial z = 10 \partial p_\infty/\partial x$ for $t>0$.}
\label{fig:sspg_twx_ret1000}
\end{figure}

The streamwise wall stress response to the SSPG is shown in figure \ref{fig:sspg_twx_ret1000}. As can be seen, the wall model is unable to capture the slight initial decrease in $\tau_{wx}$ due to a complex three-dimensional mechanism discussed in \cite{lozano2020non}. The reason is that the scrambling of momentum transporting turbulent structures due to the sudden spanwise pressure gradient is not included in any part of the present model. Significantly more sophisticated modeling of the eddy viscosity in the RANS model used to derive the LaRTE equation would be required. In this case, however, the difference is less than 1-2\% of $u_{\tau0}$. The model correctly relaxes towards the DNS trend for $tu_{\tau 0}/h>1$. The increase in $\tau_{wx}$ may be attributed to the increase in Reynolds number as the  mean pressure gradient increases in magnitude even as its alignment rotates away from the x-axis.
This behavior is ``slow'' and thus is expected to be captured well by the LaRTE model.  After a long time, the new equilibrium condition is reached in which the x-component of the pressure gradient must be balanced by the wall stress and thus the wall stress reduces back to unity as shown in 
figure \ref{fig:sspg_twx_ret1000}(b).

\begin{figure}
\centering
\includegraphics[width=0.9\linewidth]{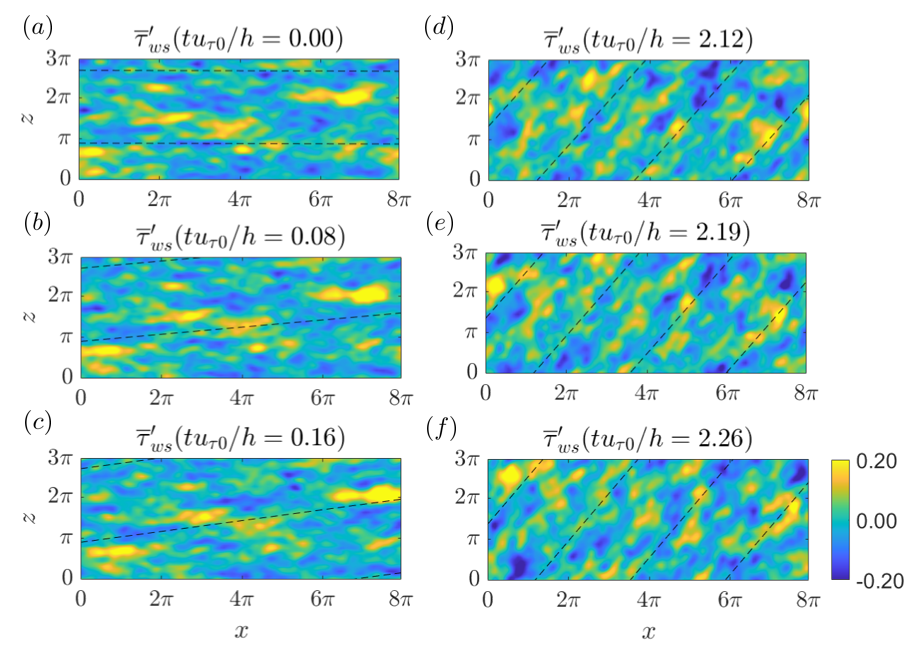}
\caption{Contours of the quasi-equilibrium s-component wall stress ($\bm{s}$ introduced in \S \ref{sec:LaRTEmodel}) with the plane-averaged mean subtracted, $\overline{\tau}_{ws}'$, for various times after the SSPG. (a-c): immediately after SSPG; (d-f): later times after SSPG. Dashed lines are aligned with the plane-averaged total wall stress angle (includes both quasi-equilibrium and non-equilibrium components).}
\label{fig:twsbar_prime_sppg}
\end{figure}

Figure \ref{fig:twsbar_prime_sppg} shows contours of fluctuations of the quasi-equilibrium  stress $\overline{\bm \tau}_w$. Specifically we show contours of 
$\overline{\tau}_{ws}'=\overline{\bm \tau}_w \cdot \langle {\bm s} \rangle - \langle \overline{\bm \tau}_w  \rangle \cdot \langle {\bm s} \rangle
$, where $\langle {\bm s} \rangle$ is the plane averaged unit vector, i.e. in the direction of the mean LaRTE wall stress.  
The contours represent the wall stress fluctuations aligned with the plane-averaged mean quasi-equilibrium wall stress direction. As a reference, the dashed lines shown are aligned with the total wall stress and thus include the contributions from the laminar boundary layer developing due to the application of the SSPG. The application of the SSPG disrupts the orientation and shape of the structures as the mean flow rotates towards the $z$ direction. At later times after the SSPG is applied, subfigures (d-f) show that the structures have had enough time to orient and advect themselves with the mean wall stress direction, albeit with a reduced size. 

\begin{figure}
\centering
\includegraphics[width=0.5\linewidth]{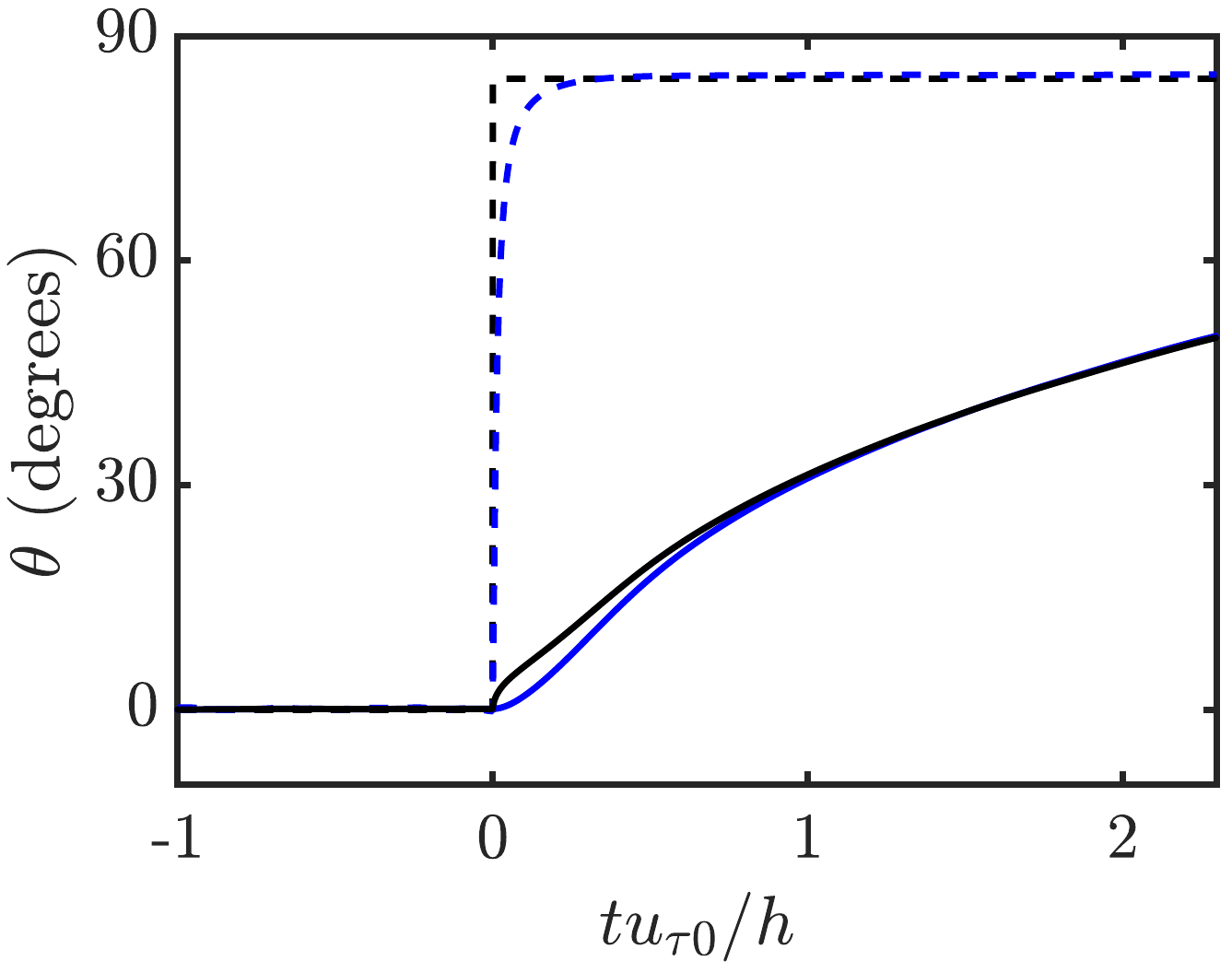}
\caption{Plane-averaged pressure gradient and wall stress angles after the SSPG. $\langle \bm{\nabla}_h \overline{p} \rangle$ (\protect\thickbluedashed), $\langle \bm{\nabla}_h p_\infty \rangle$ (\protect\thickblackdashed),
$\langle \overline{\bm{\tau}}_w \rangle$ (\protect\thickblueline),
$\langle \overline{\bm{\tau}}_w \rangle + \langle \bm{\tau}_w'' \rangle$ (\protect\thickblackline) where angled brackets indicate plane averaging.}
\label{fig:angles_sspg}
\end{figure}

Figure \ref{fig:angles_sspg} shows 
the evolution of the angle (with respect to the $x$ axis) of the pressure gradients (dashed lines) and resuling plane averaged total (black line) and quasi-equilibrium (blue line) wall stress. As expected, the quasi-equilibrium component of the wall stress has a delayed and smoothed response to the SSPG, while at the initial transient the non-equilibrium component dominates the plane averaged wall stress response.  After some adjustment time the quasi-equilibrium component becomes more dominant, again in establishing the trends of the plane averaged wall stress components such as its direction. At large times, the angle tends to the same angle as the net applied pressure gradient.

\section{Summary and conclusions}
\label{sec:conclusions}
We have introduced the Lagrangian Relaxation Towards Equilibrium (LaRTE) wall model, representing quasi-equilibrium dynamics. The LaRTE model consists of an evolution equation for the friction velocity (and thus the wall stress) by using a method similar to that introduced in \cite{chung2009large} where the law of the wall is utilized to rewrite the unsteady term as $\partial \bm{u}_\tau /\partial t$. Also similar to \cite{chung2009large} and \cite{yang2015integral}, the LaRTE model is based on the vertically integrated RANS-like thin boundary layer equations. Remarkably, it is found that generalization to include the advection terms leads to a Lagrangian form rather than an Eulerian one. Moreover, the right-hand side of this equation is in the form of a  term  describing relaxation towards the stress at the wall model height, with a relaxation time scale $T_s$. A-priori testing based on DNS channel flow data shows that the relaxation time scale is consistent,
whereas use of shorter time scales would be inconsistent, with the assumption of quasi-equilibrium.

The proposed formalism  allows for separate modeling of non-equilibrium effects not captured by the LaRTE model. In \S \ref{sec:stokeslayer} we introduce a non-equilibrium wall model to capture quick transient pressure gradient effects. The approach is well suited for modeling the laminar Stokes layer observed in the literature for flows with a rapidly changing pressure gradient.

The LaRTE plus laminar non-equilibrium model is tested for (a) simple channel flow with a constant pressure gradient and (b) the sudden spanwise pressure gradient test case introduced in \cite{lozano2020non}. These two cases were selected in order to verify that the LaRTE model elements indeed exhibit quasi-equilibrium behaviour, develop intuition regarding the model performance, and examine its response in a case where the non-equilibrium effects are known to follow the Stokes laminar solution. 
Time signals, wall stress contour plots, and wall stress PDFs reveal interesting and useful physical insight as to how the model operates. 
For example, time signals of the LaRTE-predicted wall stress are consistent with the idea that the friction velocity relaxes towards its equilibrium value at the relaxation time scale. 
Wall stress contour plots show that the   structures in the LaRTE  model are rather large, of length on the order of or larger than the channel half-width. LaRTE thus implicitly averages out the turbulence leaving only large-scale fluctuations that are internally consistent with ``quasi-equilibrium'', i.e. under conditions that are sufficiently averaged so that using the locally determined friction velocity one may assume the law of the wall to hold. 

A comparison between Lagrangian versus Eulerian versions of the relaxation towards equilibrium model show the Lagrangian version tends to advect wall stress structures whereas the Eulerian version tends to thin out and elongate structures in the streamwise direction. Wall stress PDFs show the distribution of wall stress fluctuations of each component and how the non-equilibrium model diminishes in importance as Reynolds number increases. Differences between PDFs from the wall model and filtered DNS show that additional ingredients will be required to fully capture the turbulent fluctuations in wall stress, especially in the streamwise direction. For the sudden spanwise pressure gradient test case, time signals of the plane-averaged mean spanwise and streamwise wall stress show the model is in good agreement with the DNS data of \cite{lozano2020non} and that long after the application of the SSPG the model approaches the correct steady state values.

In the LaRTE approach,   closure for the total stress $\overline{\bm \tau}_\Delta$ at the wall-model height is required. Also, the pressure gradient decomposition into various time-scales involves some modeling choices. 
While a number of other options exist and could be explored, the corresponding choices used in this paper were justified through physical interpretation and the model's ability to yield good results for the test cases considered here. 

In summary, the proposed LaRTE model represents a new framework for wall modeling.  Specifically, the formal identification of quasi-equilibrium dynamics enables us to model the remainder non-equilibrium parts more rigorously. Moreover, the quasi-equilibrium portion can be further expanded beyond the case of smooth flat-plate surfaces to include, e.g.,  roughness effects and additional dependencies on other physical  parameters such as streamline curvature, thermal and compressibility effects, etc. The choice of model for the turbulence stress at the wall-model height and the pressure gradient decompositions could also be further improved. We leave such extensions for future work.  Also, the new wall model must be tested in other flows such as flows with streamwise pressure gradients,  flow over a cylinder, steps, and others. 
Finally, for flows with rapidly oscillating walls for e.g. drag reduction \citep{jung_mangiavacchi_akhavan_1992} in which the turbulent structure below $y=\Delta$ is heavily affected by non-equilibrium scrambling effects, additional modeling will be required, corresponding to the ``white'' region in figure \ref{fig:scale_schematic}.

 \vspace{0.2in}
 
\noindent {\bf Acknowledgements:} The authors are grateful to  Dr.  Shidong  Jiang for providing us with the code to compute the SOE parameters.  

 \vspace{0.1in}
\noindent {\bf Funding:} Funding was provided by the Office of Naval Research, grants N00014-17-1-2937 and N00014-21-1-2162. The program manager was Dr. Peter Chang. 

 \vspace{0.1in}
\noindent {\bf Declaration of interests:} The authors report no conflict of interest.

 \vspace{0.1in}
\noindent {\bf Author ORCIDs:}
T. A. Zaki https://orcid.org/0000-0002-1979-7748

C. Meneveau https://orcid.or/0000-0001-6947-3605 

M. Fowler
https://orcid.org/0000-0001-6839-442X

  \vspace{0.1in}
  
\noindent  {\bf Author contributions.} All three authors contributed to deriving theory, reaching conclusions, and writing the paper. M.F.  performed the simulations and analysis of simulation data.

\appendix

\section{Derivation of integral of advective term}
\label{app:advection}
 Here we provide evaluation of the vertical integration of the advective terms in equation \ref{eq:horizmom},
 namely 
 \be 
\int_0^\Delta \left[ \partial_s (\overline{u}_s \, \overline{\bm u}_s) + \partial_y(\overline{v} \, \overline{\bm u}_s) \right] \,dy.
\ee
We develop the second term 
$\int_0^\Delta \partial_y(\overline{v} \, \overline{  u}_s {\bm s}) \,dy = \overline{v}(\Delta) \overline{u}_s(\Delta) \,\,{\bm s},$
which together with equation \ref{eq:vertvel} can be written as
 \be
\int_0^\Delta \partial_y(\overline{v} \, \overline{  u}_s {\bm s}) \,dy = \overline{v}(\Delta) \overline{u}_s(\Delta) \,\,{\bm s} = - \frac{\partial u_\tau}{\partial s} \Delta f^2(\Delta^+) (u_\tau {\bm s}).
\label{eq:vertterm}
\ee
Next, we develop the term $\int_0^\Delta   \partial_s (\overline{u}_s^2 \, {\bm s}) \, dy$ that can be written as 
  \be
 \frac{\partial}{\partial s} \left( u_\tau^2  \, {\bm s} \,  \int_0^\Delta f^2(y^+) \, dy \right) = (u_\tau   \, {\bm s} )\,  \frac{\partial}{\partial s}  \left( u_\tau  \int_0^\Delta f^2(y^+) \, dy \right) + \left( u_\tau\int_0^\Delta f^2(y^+) \, dy \right) \frac{\partial}{\partial s}(u_\tau  \, {\bm s} ).
  \ee 
 Using 
 \be 
 \frac{\partial}{\partial s} \left[  u_\tau \int_0^\Delta f^2(y^+) \,dy \, \right] = 
  \frac{\partial u_\tau}{\partial s}  \int_0^\Delta \left[ f^2(y^+) + y^+ d f^2  /dy^+ \right] dy = \frac{\partial u_\tau}{\partial s} \Delta f^2(\Delta^+),
 \ee 
 we obtain
 \be 
 \int_0^\Delta   \partial_s (\overline{u}_s^2 \, {\bm s}) \, dy = \left( u_\tau\int_0^\Delta f^2(y^+) \, dy \right) \frac{\partial}{\partial s}(u_\tau  \, {\bm s} )  + \frac{\partial u_\tau}{\partial s} \Delta f^2(\Delta^+) (u_\tau {\bm s}).
 \ee 
 Combining with the vertical advection term evaluated in equation \ref{eq:vertterm} we see that the last term cancels exactly and we obtain simply
 \be 
\int_0^\Delta \left[ \partial_s (\overline{u}_s^2 \, {\bm s}) + \partial_y(\overline{v} \, \overline{\bm u}_s) \right] \,dy  =  \left( u_\tau\int_0^\Delta f^2(y^+) \, dy \right) \frac{\partial}{\partial s}(u_\tau  \, {\bm s} ) = \int_0^\Delta f^2(y^+) \, dy \,\,\, {\bm u}_\tau \cdot \bm{\nabla}_h {\bm u}_\tau
\ee
since $u_\tau {\bm s} = {\bm u}_\tau$ 
and $ \partial_s = {\bm s} \cdot \bm{\nabla}_h$. Furthermore, from the definitions of the cell thicknesses $\delta^*_\Delta$ and $\theta_\Delta$ it is easy to show that
\be 
\int_0^\Delta f^2(y^+) \, dy = f^2(\Delta^+) \Delta \left(1-\frac{\delta^*_\Delta}{\Delta} - 
\frac{\theta_\Delta}{\Delta}\right). 
\ee 
 Division by $\Delta f(\Delta^+)$ and multiplication by $u_\tau {\bm s}={\bm u}_\tau$ leads to the advective velocity ${\bm V}_\tau$ as stated in equation \ref{eq:defVtau}.

 \section{Derivation of integral of horizontal diffusion term}
 \label{app:diffusion}
 Here we develop one of the expressions needed for the diffusive term, namely
 \be {\cal \bm D}_{\tau\nu} = \int_0^\Delta   \nu \left[  \bm{\nabla}_h \overline{\bm{u}}_s+(\bm{\nabla}_h \overline{\bm{u}}_s)^\top \right]   \,  dy,
\ee
i.e. the contribution to ${\cal \bm D}_{\tau}={\cal \bm D}_{\tau\nu}+{\cal \bm D}_{\tau T}$  from the constant (molecular) viscosity.  Using again $ \overline{\bm u}_s = \overline{u}_s(y)\, {\bm s}$ and $\bm{\nabla}_h = {\bm s}\,\partial_s$, we consider
 \be 
 \nu \int_0^\Delta {\bm s} \,\, \partial_s \left[ \overline{u}_s(y) {\bm s} \right] \, dy = 
 \nu  \int_0^\Delta \partial_s \left[  u_\tau f(y^+) \right]  \, dy \, ( {\bm s} \, {\bm s} ) \,+
 \nu \,    u_\tau \int_0^\Delta f(y^+) \, dy \,\, \left( {\bm s} 
 \, \frac{\partial {\bm s}}{\partial s} \right).
 \ee 
 Since $\int_0^\Delta \partial_s \left[  u_\tau f(y^+) \right] dy = \Delta f(\Delta^+) \, \partial u_\tau/\partial s$, we obtain
 \begin{align}
 \nu \int_0^\Delta {\bm s} \, \partial_s \left[ \overline{u}_s(y) {\bm s} \right] dy &= \nu  \Delta f(\Delta^+) \, \frac{\partial u_\tau}{\partial s} \, ( {\bm s}   {\bm s} ) + \nu u_\tau \int_0^\Delta f(y^+) \, dy \, \left( {\bm s} 
  \frac{\partial {\bm s}}{\partial s} \right) \nonumber \\
 &= \nu  \Delta f(\Delta^+) \,   {\bm s}  \, \frac{\partial (u_\tau {\bm s}) }{\partial s} - \nu u_\tau \delta^*_\Delta \left( {\bm s} 
 \, \frac{\partial {\bm s}}{\partial s} \right).
 \end{align}
 Adding the transpose and writing $s_i\partial s_j + s_j\partial s_i=\partial (s_i s_j)$ as $\partial_s({\bm s}{\bm s}) = {\bm s}\cdot {\bm \nabla}_h({\bm s}{\bm s})$ we obtain, finally,
 \be {\cal \bm D}_{\tau\nu} =  \nu \, \Delta \, f(\Delta^+) \left[ \bm{\nabla}_h \bm{u}_\tau+(\bm{\nabla}_h \bm{u}_\tau)^\top \right] - \nu \,  \delta^*_\Delta \,  \,
 {\bm u}_\tau \cdot {\bm \nabla}_h\left({\bm s}\,{\bm s}\right).
\ee
 The first term is in the form of standard horizontal diffusion proportional to viscosity and the symmetric part of the  friction velocity horizontal gradient tensor. The last term is non-standard and  represents spatial direction changes. However, since $\delta^*$ is expected to be typically much smaller than $\Delta f(\Delta^+)$ this term can be expected to be small (similarly to the term in the Eulerian time derivative proportional to $\delta^*_\Delta$ in Eq. \ref{eq:dLdt}).
 
 The case of $y$-dependent eddy viscosity to determine ${\cal \bm D}_{\tau T}$ can be developed similarly but includes more complicated expressions. In the simulations presented in this paper the horizontal diffusion terms are not explicitly included while some horizontal diffusion is provided by the low-order discretization method employed to solve the Lagrangian advection part of the equation efficiently..

 \section{Fits for $f(\Delta^+)$, cell thickness scales, and $Re_{\tau \Delta}$}
 \label{app:fits}
 
Here we provide details how the cell thickness scales $\delta^*_\Delta$ and $\theta_\Delta$, and law of the wall quantities $f(\Delta^+)$ and $Re_{\tau \Delta}$, are evaluated. 
We begin with the velocity profile fit to the numerically obtained 
solution to the 
``full equilibrium'' streamwise momentum equation with a prescribed eddy viscosity. As in \cite{meneveau2020note}, a mixing length model with the van-Driest damping function is used. All variables are non-dimensionalized in inner units such that we can obtain $u^+ = f(y^+)$, i.e. the law of the wall. We then develop a fit for this velocity profile as a function of the wall-model height in inner units $\Delta^+ \equiv u_\tau \Delta/\nu$ where $u_\tau$ is the friction velocity magnitude obtained from the quasi-equilibrium wall model (its value at the previous time-step is used for explicit evaluation). 

\be \label{eq:velocity_fit}
f(\Delta^+) =
\left[\frac{1}{\kappa} \log (\kappa_2+\Delta^+) + B\right]
\left[1+\left(\kappa_1^{-1} \Delta^+\right)^{-\beta}\right]^{-\frac{1}{\beta}}.
\ee
We set $\kappa = 0.4$ and other fitting parameters are chosen to minimize the error between the fit and the numerical solution. This yields
\begin{equation*}
 B = 4.95, \; \kappa_2 = 9.753, \; \beta = 1.903, \; \kappa_1 = \frac{1}{\kappa}\log(\kappa_2)+B,
\end{equation*}
where the last choice is required to ensure the near-wall viscous 
layer asymptote $f(\Delta^+)=\Delta^+$.  The fit has an error, relative to the numerical solution, of less than one percent for $\Delta^+>5$ and an error less than $2.25\%$ for $0\leq\Delta^+\leq 5$. Other fits could also be used, such as the traditional Reichardt fit \citep{reichardt1951vollstandige}: $f(\Delta^+) = \kappa^{-1}\log(1+\kappa \Delta^+) + 7.3[1-\exp(-\Delta^+/11)-(\Delta^+/11)\exp(-\Delta^+/3)]$. But it yields over 3\% relative error near $\Delta^+ \sim 10$, so we prefer to use Eq. \ref{eq:velocity_fit} instead.

From the numerical solution of the velocity profile we can obtain the cell  thickness scales from their definitions in Eqs. \ref{delta_star_Delta_def} and \ref{theta_Delta_def} through numerical integration. We then provide fits for these  scales (non-dimensionalized by $\Delta$) to eliminate the need for numerical integration while implementing the wall model. The fits were developed by analytically determining the length scales when $\Delta$ lies in the viscous sublayer or when $\Delta$ lies in the log-layer and then using a merger function to create a function valid for any $\Delta$ within the range of $\Delta^+$ considered. The log-layer solutions are denoted with superscript ``log''. The cell displacement thickness fit is 
\be \label{delta_star_Delta_fit}
    \frac{\delta_{\Delta}^{*}}{\Delta} =
    \frac{1}{2} \gamma_1 + (1-\gamma_1)^{C_3}
    \left(\frac{\delta^{*log}_\Delta}{\Delta}\right)
\ee
where
\begin{align*}
    \frac{\delta^{*log}_\Delta}{\Delta} = \frac{C_1}{Re_\Delta}+\frac{1}{\kappa} 
    \frac{\Delta^+}{Re_{\Delta}}, \quad
    \gamma_1=\frac{1}{1+C_{2} Re_{\Delta}^{C_4}}, \\
    C_1 = 23.664, \; C_2 = 0.0016, \; C_3 = 1.516, \; 
    C_4 = 1.177
\end{align*}

\begin{figure}
\centering
\includegraphics[width=0.8\linewidth]{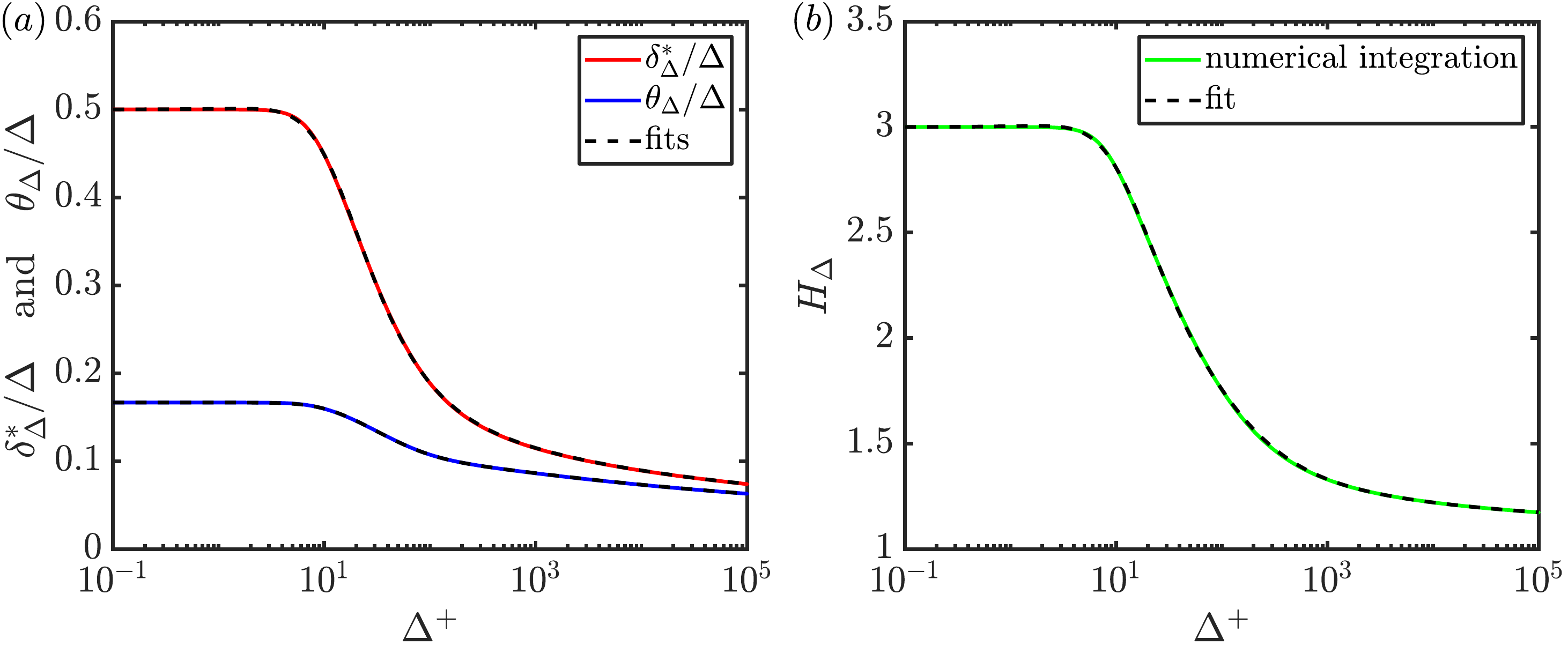}
\caption{Cell integral lengthscales (left), shape factor (right), and their fits}
\label{fig:integral_lengthscales}
\end{figure}

Similarly, the cell momentum thickness fit is
\begin{equation} \label{theta_Delta_fit}
    \frac{\theta_\Delta}{\Delta} =
    \frac{1}{6} \gamma_2 + (1-\gamma_2)^{C_8}
    \left( \frac{\theta_\Delta^{log}}{\Delta} \right)
\end{equation}
where
\begin{align*}
    \frac{\theta_\Delta^{log}}{\Delta}
    &= \frac{1}{Re_\Delta} \left(
    C_5 + \frac{\Delta^+}{\kappa} \right)
    + \frac{\Delta^+}{Re_\Delta^2}
    \left( C_6 - \frac{2 \Delta^+}{\kappa^2}
    \right), \quad
    \gamma_2=\frac{1}{1+C_7 Re_{\Delta}}, \\
    C_5 &= -103.5, \; C_6 = 2586, \; C_7 = 0.00154, \; C_8 = 2.475
\end{align*}
For both fits, $Re_\Delta = \Delta^+ f(\Delta^+)$ and $\kappa=0.4$. The fitting functions from Eqs. \ref{delta_star_Delta_fit} and \ref{theta_Delta_fit} have a maximum error of 0.5 percent over (over the range $10^{-1} \leq \Delta^+ \leq 10^5$) compared to the numerical solutions obtained by integrating the velocity profile. Note Eqs. \ref{delta_star_Delta_fit} and \ref{theta_Delta_fit} require the fit $f(\Delta^+)$ given by Eq. \ref{eq:velocity_fit}. Figure \ref{fig:integral_lengthscales} displays the results as well as the associated cell shape factor $H_\Delta = \delta_\Delta^*/\theta_\Delta$. Note that its low Reynolds number limit is not the traditional Blasius profile value, but $H_\Delta \to 3$ associated with a linear profile. 

Furthermore, to evaluate $\bm{\tau}_\Delta$ according to Eq. \ref{eq:modelfittau} in terms of $Re^{\rm pres}_{\tau \Delta}$,  we use the fitting function provided in \cite{meneveau2020note} reproduced here for completeness. Since the applications in the present paper only deal with smooth surfaces the merging with rough wall parameterizations treated in \cite{meneveau2020note} is omitted. Mean pressure gradients are included, however, and so we use $Re^{\rm pres}_{\tau \Delta}$ that generalizes $Re^{\rm fit}_{\tau \Delta}$ to include pressure gradients. The inputs are $Re_\Delta = U_{\rm LES} \Delta/\nu$ and $\psi_p = \rho^{-1}(\nabla_h P \cdot \hat{\bm e}_u) \, \Delta^3/\nu^2$, and the fitting function \cite{meneveau2020note} is provided in algorithm \ref{alg:retd_fit}.

\begin{algorithm}
\caption{$Re_{\tau\Delta}^{\rm pres}= Re_{\tau\Delta}^{\rm pres}(Re_\Delta,\psi_p)$ where $0 < Re_\Delta < 10^7$ and $|\psi_p| < 2 \times 10^7$}
$\beta_1 =[{1+0.155/Re_\Delta^{0.03}}]^{-1}, \quad
\beta_2=1.7-[ {1+36/Re_\Delta^{0.75}} ]^{-1},$\\
$\kappa=0.4, \quad \kappa_3 = 0.005, \quad \kappa_4 = \kappa_3^{\beta_1-1/2}$,\\
$Re^{\rm fit}_{\tau \Delta} = \kappa_4 Re_\Delta^{\beta_1} [1+(\kappa_3 Re_\Delta)^{-\beta_2} ]^{(\beta_1-1/2)/\beta_2}$\\
\eIf{$\psi_p<0$}
{
    $Re_{\tau\Delta-{\rm min}}(\psi_p) = 1.5 (-\psi_p)^{0.39} \left[1+\left(\frac{1000}{-\psi_p}\right)^2 \right]^{-0.055},$\\
    $p(\psi_p) = 2.5-0.6\left[1+\tanh(2(\log_{10}(-\psi_p)-6))\right],$\\
    $Re_{\tau \Delta}^{\rm pres} = \left( (Re_{\tau\Delta-{\rm min}})^{p} + (Re_{\tau\Delta}^{\rm fit})^{p} \right)^{1/p}$
}
{
    $Re_{\Delta-{\rm min}}(\psi_p) =  2.5 \psi_p^{0.54}\left(1+\left[\frac{30}{\psi_p}\right]^{1/2} \right)^{-0.88}$\\
    \eIf{$Re_\Delta > Re_{\Delta-{\rm min}}$}
    {
        $Re_{\tau \Delta}^{\rm pres}   =  Re_{\tau\Delta}^{\rm fit} \left(1-(1+\log[Re_\Delta/Re_{\Delta-{\rm min}}])^{-1.9} \right)$
    }
    {
        $Re_{\tau \Delta}^{\rm pres}   = 0$ (flow separation)
    }
}
\label{alg:retd_fit}
\end{algorithm}

\section{Verification of the accuracy of the SOE method}
\label{app:soe}

\begin{figure}
\centering
\includegraphics[width=\linewidth]{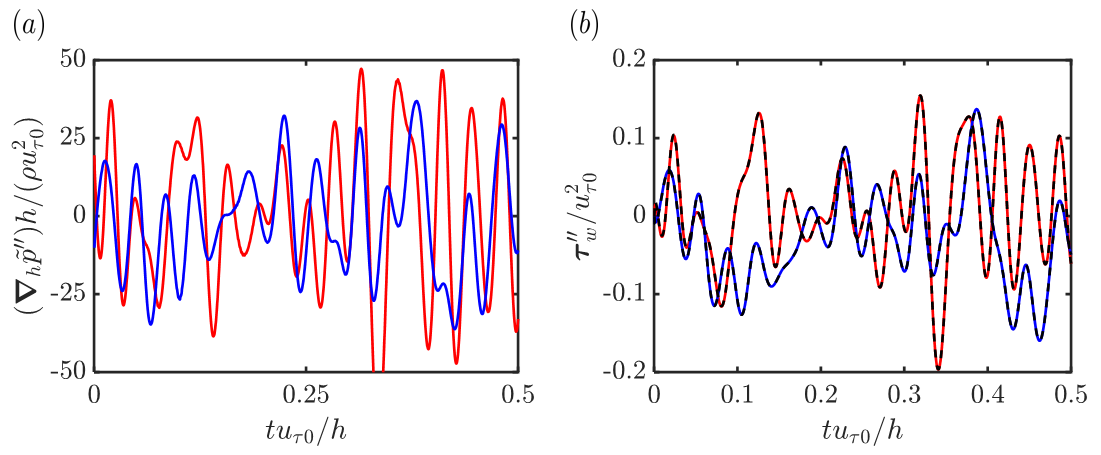}
\caption{(a) representative non-equilibrium pressure gradient signal and (b) corresponding non-equilibrium wall stress. Red and blue curves correspond with the x and z components, respectively. For (b), solid lines are computed using the L1 method and dashed black lines are computed using the SOE method. The SOE constants are computed using $\delta t=4\times 10^{-4}$, $\epsilon=10^{-9}$, and $T=1$ (yielding $N_{\exp}=48$).}
\label{fig:soe_error_analysis}
\end{figure}

We now validate the accuracy of the sum-of-exponentials (SOE) method presented in \S \ref{sec:convolution_integral} by comparing it with the more costly, but well established ``L1 method'' \citep{li_zeng_2015}. A typical pressure gradient signal from real flow simulations for $Re_\tau =1000$ and $\Delta = h/30$ is used for the comparison. The SOE constants $\omega$ and $s$ and the number of exponential terms, $N_{\exp}$, are functions of the error of the SOE approximation, $\epsilon$, the time step size, $\delta t$, and the length of time considered, $T$. The method guarantees that $|t^{-1/2} - \Sigma_{m=1}^{N_{\exp}} \omega_m \exp(-s_m t) | \leq \epsilon$ but it does not guarantee that the error in computing the integral in \ref{eq:neq_model_discrete} is below a desired value. Therefore, we simply compare the SOE method with the L1 method to show that the errors between the two are insignificant despite the significant computational cost differences between the two numerical methods. The results are presented in figure \ref{fig:soe_error_analysis}. (a) shows the input signal and (b) shows the output signal. The time step size used for computing the SOE constants matches the time step size of the pressure gradient signal, thus error in the approximation of the kernel is guaranteed. Visually, the differences between the two methods are not noticeable. Quantitatively the root-mean-sum (RMS) difference between the methods, normalized with $u_{\tau 0}^2$, is less than 0.021 in the x-direction and less than 0.011 in the y-direction. Therefore, differences between the two methods are shown to be small enough to be neglected for current applications. The L1 method requires $O(N_T)$ computations per time step and $O(N_T)$ terms to be stored. The SOE method on the other hand requires $O(N_{\exp})$ computations per time step and $(N_{\exp})$ terms to be stored where $N_{exp}$ can be estimated from equation \ref{eq:Nexp}. For a large number of time steps $N_{\exp} \sim O(\log N_T)$ which in practice is held constant to avoid a dynamic storage size. Therefore, the SOE method has been shown to be significantly cheaper (with a non-dynamic storage size) relative to the L1 method while still providing accurate results.

The SOE constants, $\omega$ and $s$, used in all simulations in this paper are included in the supplementary materials. Dr. Shidong Jiang kindly provided the source code for computing these coefficients.

 \section{Alternate form of friction velocity evolution equation}
\label{sec:LaRTE_rewritten}

At the end of \S \ref{sec:Lagtimederivativenumerics} it is mentioned that the $\partial_t \bm{s}$ term in the friction velocity evolution equation is discretized using $\bm{s}$ known at the $n-1$ and $n-2$ time steps. This is done to avoid coupling between the $x$ and $z$ evolution equations. 
Alternatively, it is possible to rewrite equation \ref{eq:Lagrelax} without an additional $\partial_t {\bm s}$ time derivative term. The resulting set of equations require additional evaluations of spatial gradients, thus we prefer working with equation \ref{eq:Lagrelax}. However, for completeness we here provide the alternate form of the LaRTE equation that does not contain the non-standard $\partial_t \bm{s}$ term.

We begin with equation \ref{eq:Lagrelax} and move all the time derivative terms to the left-hand-side (LHS) of the equal sign and the remaining terms to the right-hand-side (RHS). The resulting equation is
\begin{equation} \label{eq:LaRTE_rewrite1}
\frac{\partial  {\bm{u}}_\tau}{\partial t} 
- u_\tau \frac{\delta^*_\Delta}{\Delta} \frac{\partial {\bm s}}{\partial t}
=  {\bf {RHS}}
\end{equation}
 where
\begin{equation} \label{eq:LaRTE_rewrite_RHS}
{\bf RHS} = - {\bm V}_\tau \cdot \bm{\nabla}_h {\bm{u}}_\tau + \frac{1}{T_s} \left[ \frac{1}{u_\tau}\left(-\frac{\Delta}{\rho} \bm{\nabla}_h \overline{p} + \overline{\bm{\tau}}_\Delta \right) - {\bm u}_\tau\right]\, + \, \frac{1}{\Delta f(\Delta^+)} \,\, {\bm \nabla}_h \cdot {\cal \bm D}_\tau
\end{equation}
with the same definitions presented in \S \ref{sec:LaRTEmodel}. Then using the definition $\bm{s} \equiv \bm{u}_\tau/u_\tau$ we write $\partial_t \bm{s}$ in terms of friction velocity time derivatives, i.e.
\begin{equation}
\frac{\partial \bm s}{\partial t}
= \frac{1}{u_\tau} \frac{\partial \bm{u}_{\tau}}{\partial t}
- \frac{\bm{u}_{\tau}}{u_\tau^3} \left(
{\bm u}_\tau \cdot \frac{\partial {\bm u}_\tau}{\partial t}  \right).
\end{equation}
Using equation \ref{eq:LaRTE_rewrite1} we replace $\partial_t \bm{u}_\tau$ with ${\bf RHS}+u_\tau(\delta_\Delta^*/\Delta)\partial_t {\bm s}$ and solve for $\partial_t {\bm s}$ to get
\begin{equation}
\frac{\partial {\bm s}}{\partial t} = \left( 
1-\frac{\delta_\Delta^*}{\Delta}\right)^{-1}\left[ \frac{1}{u_\tau} {\bf RHS} - \frac{{\bm u}_\tau}{u_\tau^3} \left( {\bm u}_\tau \cdot {\bf RHS} \right) \right]
\end{equation}
where we have utilized ${\bm u}_\tau \cdot \partial_t {\bm s}=0$. 
Substituting back into equation \ref{eq:LaRTE_rewrite1} we obtain a governing equation for the friction velocity vector without the $\partial_t \bm{s}$ term:
\begin{equation}
\frac{\partial {\bm u}_\tau}{\partial t}
= {\bf RHS} - \frac{\delta_\Delta^*}{\Delta-\delta_\Delta^*}\left( {\bm s}({\bm s}\cdot {\bf RHS}) - {\bf RHS}\right).
\end{equation}
We can further simplify this equation using the vector identity
$
{\bm s} \times ({\bm s} \times {\bf RHS})
= {\bm s}({\bm s}\cdot {\bf RHS}) - {\bf RHS}({\bm s} \cdot {\bm s})$ so that the
the final, rewritten evolution equation for ${\bm u}_\tau$ is 
\begin{equation} \label{eq:LaRTE_rewrite_final}
\frac{\partial {\bm u}_{\tau}}{\partial t}+{\bm V}_\tau \cdot
{\bm \nabla}_h {\bm u}_{\tau}=\frac{1}{T_s} \left[ ... \right] 
+ \frac{1}{\Delta f(\Delta^+)} 
{\bm \nabla}_h \cdot  {\cal \bm D}_\tau -
\frac{\delta^*_\Delta}{\Delta-\delta^*_\Delta} \, {\bm s} \times ( {\bm s} \times {\bf{RHS}} ),
\end{equation}
where $...$ is the entire relaxation term in the square parenthesis in Eq. \ref{eq:LaRTE_rewrite_RHS}. Equation \ref{eq:LaRTE_rewrite_final} is an alternate form of writing the LaRTE wall model equation, again confirming the Lagrangian relaxation dynamics for ${\bm u}_\tau$ but now including the additional term proportional to $\delta^*_\Delta/(\Delta-\delta^*_\Delta)$ written without time-derivatives. Again, similarly to $\partial_t{\bm s}$, this term is perpendicular to $\bm s$. However, in seeking to implement this term using the Lagrangian time derivative approach the evaluation of ${\bf RHS}$ in the cross product becomes cumbersome because it contains the advective term ${\bm V}_\tau \cdot {\bm \nabla}_h {\bm u}_\tau$ and we would require additional evaluation of the spatial gradients of ${\bm u}_\tau$ at the upstream location. This is why equation \ref{eq:Lagrelax} is the preferred form for the LaRTE wall model, at least in our implementation.

\bibliographystyle{jfm}
\bibliography{references,meneveau-biblio}
 
\end{document}